\documentclass[12pt]{iopart}
\usepackage{iopams}
\usepackage{ifpdf}
\usepackage{amssymb}
\usepackage{txfonts}
\usepackage{graphicx}
\usepackage{subfigure}
\usepackage{dcolumn}
\usepackage{bm}
\usepackage{sidecap}
\usepackage{enumitem}
\usepackage{cite}
\usepackage{color}

\begin{document}

\title{High Speed Continuous Variable Source-Independent Quantum Random Number Generation }

\author{Bingjie Xu,$^{1,2,5}$ Ziyang Chen,$^{3,5}$ Zhengyu Li,$^{3,6}$ Jie Yang,$^{1}$ Qi Su,$^{2,6}$ Wei Huang,$^{1}$ Yichen Zhang$^{4}$ and Hong Guo$^{3}$}

\address{$1$ Science and Technology on Security Communication Laboratory, Institute of Southwestern Communication, Chengdu 610041, China}
\address{$2$ State Key Laboratory of Cryptography, Beijing 100878, China}
\address{$3$ State Key Laboratory of Advanced Optical Communication Systems and Networks, School of Electronics Engineering and Computer Science, and Center for Quantum Information Technology, Peking University, Beijing 100871, China}
\address{$4$ State Key Laboratory of Information Photonics and Optical Communications, Beijing University of Posts and Telecommunications, Beijing 100876, China}
\address{$5$ These authors contributed equally to this work.}
\address{$6$ Author to whom any correspondence should be addressed.}

\ead{lizhengyu@pku.edu.cn and suq@sklc.org}
\vspace{10pt}
\begin{indented}
\item[]November 2018
\end{indented}

\begin{abstract}
As a fundamental phenomenon in nature, randomness has a wide range of applications in the fields of science and engineering. Among different types of random number generators (RNG), quantum random number generator (QRNG) is a kind of promising RNG as it can provide provable true random numbers based on the inherent randomness of fundamental quantum processes. Nevertheless, the randomness from a QRNG can be diminished (or even destroyed) if the devices (especially the entropy source devices) are not perfect or ill-characterized. To eliminate the practical security loopholes from the source, source-independent QRNGs, which allow the source to have arbitrary and unknown dimensions, have been introduced and become one of the most important semi-device-independent QRNGs. Herein a method that enables ultra-fast unpredictable quantum random number generation from quadrature fluctuations of quantum optical field without any assumptions on the input states is proposed. Particularly, to estimate a lower bound on the extractable randomness that is independent from side information held by an eavesdropper, a new security analysis framework is established based on the extremality of Gaussian states, which can be easily extended to design and analyze new semi-device-independent continuous variable QRNG protocols. Moreover, the practical imperfections of the QRNG including the effects of excess noise, finite sampling range, finite resolution and asymmetric conjugate quadratures are taken into account and quantitatively analyzed. Finally, the proposed method is experimentally demonstrated to obtain high secure random number generation rates of $15.07$ Gbits/s in off-line configuration and can potentially achieve $6$ Gbits/s by real-time post-processing.
\end{abstract}

%
%
%
%
%

\section{Introduction}

Random numbers are of extreme importance for a wide range of applications in both scientific and commercial fields~\cite{PRNG}, such as numerical simulations, lottery games and cryptography. A significant example is the quantum key distribution (QKD), in which the true random numbers are essential for both quantum states preparation and detection to guarantee unconditional security~\cite{QKD1,QKD2,QKD3}. Classical pseudo random number generators (PRNG), which are based on the computational algorithms, have been widely used in modern information systems. However, due to the deterministic and thus predictable features of the algorithms, PRNG are not suitable for certain applications where true randomness is required. Distinct from the PRNG, true random number generators (TRNG) extract randomness from physical random processes~\cite{TRNG}. An important type of TRNGs is the quantum random number generator (QRNG), which is based on the intrinsic randomness of fundamental quantum processes and can provide truly unpredictable and irreproducible random numbers~\cite{Rev1,Rev2,Rev3}.

The existing QRNG protocols can be mainly classified into three different categories as in Ref.~\cite{Rev2}, i.e. the practical, device-independent and semi-device-independent QRNGs. Till now, various practical QRNG protocols, which can realize a high random number generation rate with relatively low cost~\cite{Rev2}, have been proposed and demonstrated, including measuring photon path~\cite{QRNG_SP1,QRNG_SP2}, photon arrival time~\cite{QRNG_Arr1,QRNG_Arr2,QRNG_Arr3,QRNG_Arr4,QRNG_Arr5}, photon number distribution~\cite{QRNG_PND1,QRNG_PND2,QRNG_PND3,QRNG_PND4}, vacuum fluctuation~\cite{QRNG_Vac1,QRNG_Vac2,QRNG_Vac3,QRNG_Vac4,QRNG_Vac5,QRNG_Vac7,QRNG_Vac6}, phase noise~\cite{QRNG_Phase1,QRNG_Phase2,QRNG_Phase3,QRNG_Phase4,QRNG_Phase5,QRNG_Phase6,QRNG_Phase7,QRNG_Phase8,QRNG_Phase9,QRNG_Phase10,QRNG_Phase11} and amplified spontaneous emission noise~\cite{QRNG_ASE1,QRNG_ASE2,QRNG_ASE3,QRNG_ASE4,QRNG_ASE5} of quantum states. However, practical QRNGs can produce true random numbers with information-theoretical provable security only if the randomness source and detection devices are trusted and fulfill with the model assumptions, which usually fails in cases that the devices are complex or controlled by eavesdroppers~\cite{Rev2}.
To avoid the defects, device-independent (DI) QRNG, which can generate verifiable randomness without assumptions on the source and measurement devices by observing the violation of Bell's inequality, have been proposed~\cite{Bell,QRNG_DI1}. Although DI-QRNG protocols (including both randomness expansion~\cite{QRNG_DI2,QRNG_DI3,QRNG_DI4,QRNG_DI5} and amplification~\cite{QRNG_DI6,QRNG_DI7} protocols) have advantage of the self-testing randomness, they are highly challenging in realistic implementations (e.g. not loss tolerant), and the generation speed is usually too slow for practical applications~\cite{QRNG_DI5}. Thus, QRNG protocol with reasonable assumptions and high practical performance is meaningful and greatly needed. To balance the performance and the security, the semi-device-independent QRNG provides a trade-off between the practical and device independent QRNGs, where high speed and low-cost informational provable randomness can be generated under several reasonable assumptions without requiring trusted and complete model assumptions on all devices~\cite{Rev2}. In general, there are two types of semi-device-independent QRNGs, namely, measurement-device-independent (MDI) QRNG~\cite{QRNG_SDI4,QRNG_SDI5,QRNG_SDI3} and source-independent (SI) QRNG~\cite{QRNG_SDI2014, QRNG_SDI1, QRNG_SDI2, QRNG_SDI_HET}. MDI-QRNG scenarios require untrusted measurement devices, whereas the source needs to be
well characterised. As a contrast, SI-QRNGs always assume that the entropy source is unknown to users and thus totally untrusted, followed by a well modeled measurement device. Particularly, as the quantum entropy source device is usually a complicated physical system in practice and crucial for randomness generation, any deviations in the real-life implementation from its model assumptions may affect the output randomness~\cite{QRNG_SDI1}. Therefore, how to design a QRNG protocol without any assumption on the source device become an important and meaningful issue. To solve this problem, the first ideal of generating randomness with unknown source was proposed in Ref.~\cite{QRNG_SDI2014}, which removes all the assumptions on the source and the dimension of Hilbert space. Then the SI-QRNG protocols exploiting discrete-variable method (based on measuring single photon path)~\cite{QRNG_SDI1} and continuous-variable (CV) methods (based on vacuum fluctuation)~\cite{QRNG_SDI2, QRNG_SDI_HET} were proposed and experimentally verified, respectively, in which the output randomness can be certified even when the source is completely uncharacterized and untrusted. These different protocols make SI protocols competitive in implementations.

It is worth noting that, because of a totally unknown source employed in SI-QRNG scenarios, a well-modeled detector is of significant importance. Recently, a model of coherent detection to quantify randomness in a full quantum scenario is investigated~\cite{coherent_detection}. This technique can also be used to analyse the SI-QRNG protocols.

In this paper, a CV-SI-QRNG protocol, which requires no assumption on the source, is proposed by measuring quadrature fluctuations of quantum optical field and experimentally demonstrated. Compared to previous works where security analysis is based on entropic uncertainty~\cite{QRNG_SDI2}, we demonstrate a new method to estimate a lower bound on the extractable randomness independent from classical or quantum side information held by an eavesdropper (Eve) based on the extremality of Gaussian states~\cite{Gaussian1}. The security analysis model shows similarity to that of CV-QKD~\cite{Gaussian2}, where rich theoretical tools exist, and can be easily extended to design and analyze new semi-device-independent CV-QRNG protocols, e.g. semi-device-independent CV-QRNG protocols with different input states or measurements. Furthermore, several practical issues of the protocol, including excess noise, finite detection range and resolution are quantitively analyzed, and the optimal choices of experimental parameters are discussed. It is shown that the proposed protocol is significantly resistant to noise and loss, which can be realized with off-the-shelf commercial devices and enable ultra-fast randomness generation rates. The final experimental secure random number generation rates reach up to 15 Gbits/s in off-line configuration based on the proposed method, and have potential to achieve $6$ Gbits/s by real-time post-processing.

\section{CV-SI-QRNG Protocol}

A schematic of the proposed CV-SI-QRNG protocol is described as follows (as shown in Fig.~\ref{fig:Setup}):
\begin{figure}[t]
\begin{center}
\includegraphics[width=0.8 \textwidth]{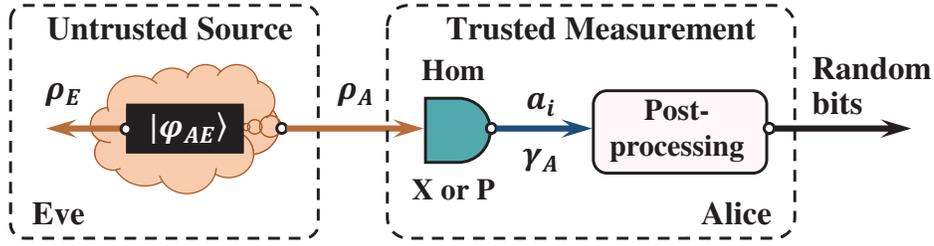}
\end{center}
\caption{(color online) Schematic of the proposed CV-SI-QRNG protocol.}
\label{fig:Setup}
\end{figure}

(i) Randomness Source Preparation: Eve prepares an untrusted and uncharacterized source with quantum states $\rho _A$ in arbitrary dimension, where Eve has access to a quantum system $E$ correlated with system $A$, and sends $\rho _A$ to the balanced homodyne measurement device of Alice.

(ii) Measurement: In one-shot experiment, Alice randomly measures the $X$ quadrature or the $P$ quadrature of $\rho _A$ to generate random bits, or randomly measure the $X$ or the $P$ quadrature to check the purity of the input states, which is based on a random initial seed. The experiments are repeated in an independently and identically distributed (i.i.d.) way, until $n_{tot}$, the number of measurement results $a_i$, are sufficient. We use $n_c$ to represent the number of check samples out of total measurements, and use $t$ to denote the length of the random seed. The measurement is assumed to be trusted and well calibrated, and all the excess noise is assumed to be introduced by a quantum correlated Eve as in Refs.~\cite{QRNG_SDI2,Gaussian2}.

(iii) Parameter Estimation: The covariance matrix (CM) $\gamma_A$ of $\rho _A$  is estimated based on Alice's measurement results on quadratures $X$ and $P$,
\begin{equation}
\gamma _A = \left[ {\begin{array}{*{20}{c}}
{{V_x}}&c\\
c&{{V_p}}
\end{array}} \right],
\end{equation}
where $V_x$ and $V_p$ are the variance of $X$ and $P$ quadratures, and $c$ is the co-variance between $X$ and $P$ quadratures for $\rho_A$.

(iv) Randomness Estimation and Extraction: Alice can extract asymptotically $({n_{tot}} - {n_c})\left( {H({a_i}) - S({a_i}:E)} \right) - t$ final randomness in $n_{tot}$ measurements by using informational provable randomness extractor, such as Toeplitz-matrix hashing extractor, where $H\left( {{a_i}} \right)$ is the Shannon entropy of Alice's measurement results $a_i$ and $S\left( {{a_i}:E} \right)$ is the quantum mutual information between Eve's quantum state $\rho_E$ and Alice's measurement results $a_i$.

In this protocol, no assumptions are made on the dimensions and purity of the input states~\cite{QRNG_SDI2}, which are difficult to verify experimentally. Indeed, it is typically difficult to prepare and keep a real quantum system in a pure state. The detection is assumed to be trusted and all the excess noise is due to a quantum correlated Eve, which shows similarity with security analysis of CV-QKD and is the most conservative option~\cite{Gaussian2}. In fact, the method can also take into account the classical side information hold by Eve effectively as in Refs.~\cite{QRNG_Vac4,QRNG_SDI2}.

\section{Security Analysis}
Suppose Alice's homodyne detection is ideal (with infinite range and resolution), then the measurement results of quadrature $X$ for $\rho_A$  is a continuous variable $a$. In the case of i.i.d. assumption, the key rate in QKD is described by the Devetak-Winter formula~\cite{D_W_Relation}, given by

\begin{equation}
K = \beta \left( {I\left( {a:b} \right) - S\left( {a:E} \right)} \right),
\end{equation}
where $\beta$ is the reconciliation efficiency, $I(a:b)$ is the classical mutual information between Alice's and Bob's data, and $S\left( {a:E} \right)$ is the Holevo's bound between Eve's quantum state $\rho_E$ and Alice's measurement results $a$, which is an upper bound of Eve's quantum side information. As a comparison, there are two different parts in QRNG scenarios comparing with that of QKD scenarios. First, Alice and Bob are at the same station locally, namely the ``source station'', and second, there is no need to perform the information reconciliation. The former leads to the reduction of the mutual information $I(a:b)$ to the discrete Shannon entropy (due to the fact that Alice actually gets a discrete variable $a_i$ in practical experiments), and the later results in the reconciliation efficiency $\beta=1$.

Note that in practice, Alice's homodyne measurement is coarse-grained with imperfect characteristics (e.g. finite range and resolution), thereby always modeled as the ideal homodyne detection together with an analog-to-digital converter (ADC) with finite sampling range, and thus the practical measurement procedure can be described as follows. First, Alice utilizes an ideal homodyne detector to measure the quadrature of input states with continuous output $a$ following probability density distribution $p(a)$, which cannot be read. Second, Alice digitizes the continuous data $a$ into $n$ bits $a_i$ following probability distribution $p_{dis}(a_i)$ by an ADC with sampling range $[ - N + \Delta /2,N - \Delta 3/2]$ and resolution $n$ (see Appendix A for details)~\cite{QRNG_Vac4}, which is the actual output of a real-life homodyne detector. Upon measurement, the continuous signal $a$ is discretized into $a_i$ over $2^n$ bins with precision $\Delta  = N/{2^{n - 1}}$.

Under the condition of practical measurements, it is easy to obtain the asymptotically extractable randomness per measurement that uniform and uncorrelated from quantum side information held by Eve, given by,
\begin{equation}
 {R_{dis}}({a_i}|E) = H\left( {{a_i}} \right) - S\left( {{a_i}:E} \right),
\end{equation}
where $H({a_i})$ is the Shannon entropy of discrete variable $a_i$, and $S\left( {{a_i}:E} \right)$ is the Holevo's bound between Eve's quantum state $\rho_E$ and Alice's measurement results $a_i$. $H(a_i)$ can be calculated easily based on Alice's measurement results $a_i$, while there are no direct way to compute $S\left( {{a_i}:E} \right)$.

To get a lower bound for the extractable randomness, one needs to upper bound $S\left( {{a_i}:E} \right)$. Fortunately, one can prove that
 \begin{equation}
 S\left( {{a_i}:E} \right)\le S\left( {a:E} \right),
 \end{equation}
which indicates that local operation on one part of a state cannot increase the mutual information between two parties (see Appendix B for detailed proofs).
For further simplifying the estimation of the upper bound of the Holevo's information $S\left( {{a}:E} \right)$, we assume Eve holds a purification of the input state, namely, ${\rho _E} = \rm{Tr}_{A}\{ \left| {{\varphi _{AE}}} \right\rangle \left\langle {{\varphi _{AE}}} \right|\}$, which is optimal for Eve. In that case, ${\rho _{AE}}$ is a pure state, so the relations $S\left( {{\rho _E}} \right) = S\left( {{\rho _A}} \right)$ and $S\left( {{\rho _E}|a} \right) = 0$ hold. Then one has,
\begin{equation}
S\left({a:E} \right) = S\left( {{\rho _E}} \right) - \int {p\left( a \right)S\left( {{\rho _{E|a}}} \right)da = S\left( {{\rho _A}} \right)},
\end{equation}
where $p(a)$ is the probability density distribution of Alice's measurement results, and $\rho_{E|a}$ is the quantum state held by Eve given that  Alice's measurement result is $a$.

If the CM of $\rho_A$  is known, one can prove that the von Neumann entropy of an arbitrary $\rho_A$  is upper bounded by that of a Gaussian state  $\rho^G_A$  with the same CM based on the extremality of Gaussian states~\cite{Gaussian1}, which means $ S\left( {{\rho _A}} \right) \le S\left( {\rho _A^G} \right)$. Alternatively, one can verify that $R$ fits all the three conditions for Lemma 1 in Ref.~\cite{Gaussian1}, which also infers that $R\left( {{\rho _A}} \right) \ge R\left( {\rho _A^G} \right)$ given a finite CM of $\rho_A$. In fact, the security of CV-QKD protocols against collective attacks have been proved with similar methods~\cite{Gaussian2}. Consequently, the lower bound of the extractable randomness can be estimated based on the following relation:
\begin{equation}
{R_{dis}}({a_i}|E) \ge H\left( {{a_i}} \right) - S\left( {a:E} \right) \ge H\left( {{a_i}} \right) - S\left( {\rho _A^G} \right).
\end{equation}
The remaining question is how to upper bound  $S\left( {\rho _A^G} \right)$ given Alice's digitized measurement results $a_i$. Due to the digitization process, we lose the information about the distribution of $a$ inside the discrete bins and outside the sampling range, thus one cannot calculate the exact CM for $\rho_A$ based on $a_i$. However, given each $a_i$ corresponding to each continuous $a$ with known upper and lower bound, one can estimate an upper bound  ${\overline V_x}\ ({\overline V_p})$ for the variance  of  $X\ (P)$ quadrature for  $\rho_A$ with a simple strategy (see Appendix C for details). Then,
\begin{equation}
S(a_i:E) \le S(\rho _A^G) \le \frac{{\overline \lambda   + 1}}{2}{\log _2}\frac{{\overline \lambda   + 1}}{2} - \frac{{\overline \lambda   - 1}}{2}{\log _2}\frac{{\overline \lambda   - 1}}{2},
\end{equation}
where $\overline \lambda   =  \sqrt {{\overline V_x}{\overline V_p}} $, and we set $c=0$ to get the upper bound of $\overline \lambda$.

In our protocol, the input state is expected (by Alice) to be a vacuum state if not disturbed by Eve, while in fact it could be an arbitrary quantum state $\rho_A$ prepared by Eve (e. g. thermal state or squeezed state). Define the variance of vacuum fluctuation as $\sigma _{{\rm{vac}}}^2 = 1$ (all the relevant quantities are normalized by vacuum fluctuation in the following). In practice, the measurement results of quantum state are unavoidable mixed with excess noise $\varepsilon$ (due to classical or quantum side information held by Eve), which is the difference between measured quadrature variance ($\sigma^2$) and the expected vacuum fluctuation, i.e. $\varepsilon=\sigma^2-1$. Define the QCNR (quantum to classical noise ratio) as $10{\log _{10}}(1/\varepsilon )$.
A typical example is that $\sigma^2=V_x=V_p=1+\varepsilon$, due to a classical Eve that controls symmetric electronic noise of the detection~\cite{QRNG_Vac4} or a quantum Eve that holds a purification of input state~\cite{QRNG_SDI2}. In this case, Alice's homodyne measurement results $a$ on quadrature $X$ (or $P$) is expected to be a continuous variable following Gaussian distribution with variance $\sigma ^2$  and null mean value,
\begin{equation}
p\left( a \right) = \frac{1}{{\sqrt {2\pi } \sigma }}exp\left( { - \frac{{{a^2}}}{{2{\sigma ^2}}}} \right).
\end{equation}
The corresponding CM of $\rho_A$  is
${\gamma _A} = \left[ {\begin{array}{*{20}{c}}
{1 + \varepsilon }&0\\
0&{1 + \varepsilon }
\end{array}} \right],$
with upper bounded
\begin{equation}
S(a:E) \le S\left( {\rho _A^G} \right) = \left( {\frac{\varepsilon} {2} + 1} \right)\log \left( {\frac{\varepsilon} {2} + 1} \right) - \frac{\varepsilon} {2}\log \frac{\varepsilon} {2}.
\end{equation}
Then, the corresponding probability distribution $p_{dis}(a_i)$ after discrete sampling is
\begin{equation}
p_{dis}({a_i}) = \left\{ \begin{array}{l}
\begin{array}{*{20}{c}}
{\frac{1}{2}erfc(\frac{{N - 0.5\Delta }}{{\sqrt 2 \sigma }}),}&{i = {i_{\min }}}
\end{array}\\
\begin{array}{*{20}{c}}
{\frac{1}{2}erf(\frac{{i + 0.5}}{{\sqrt 2 \sigma }}\Delta ) - \frac{1}{2}erf(\frac{{i - 0.5}}{{\sqrt 2 \sigma }}\Delta ),}&{{i_{\min }} < i < {i_{\max }}}
\end{array}\\
\begin{array}{*{20}{c}}
{\frac{1}{2}erfc(\frac{{N - 1.5\Delta }}{{\sqrt 2 \sigma }}),}&{i = {i_{\max }}}
\end{array}
\end{array} \right.
\end{equation}
with $i_{\min}=-2^{n-1}$ and $i_{\max}=2^{n-1}-1$. Then one can estimate the upper bounds of $V_x$, $V_p$ and $S(a_i:E)$ as in Appendix C.
Note that the model in Eqs. (4) and (5) also fit with the asymmetric quantum states (i.e. $V_x\neq V_p$).

In the above security analysis model, no assumptions are made on the input states, which remove Eve's side-information on source. Therefore, the extracted randomness is source device loophole-free.

\section{Practical issues and Numerical Simulations}
The practical issues of the proposed QRNG protocol, such as sampling range $N$, precision $n$ of homodyne detection, excess noise $\varepsilon$ and the squeeze factor $r$ of the quantum state will directly affect the performance of the protocol. Roughly speaking, the performance of the protocol attains near optimal by setting $N\in[3\sigma ,5\sigma]$ given fixed $n$ and $\varepsilon$, increases (decreases) with $n$ ($\varepsilon$) given fixed $N$, and shows resistant to excess noise.

\subsection{Effects of finite sampling range}

\begin{figure}[ht]
\centering
\subfigure[ ]{ \label{fig:subfig:a} 
\includegraphics[width=0.3 \textwidth]{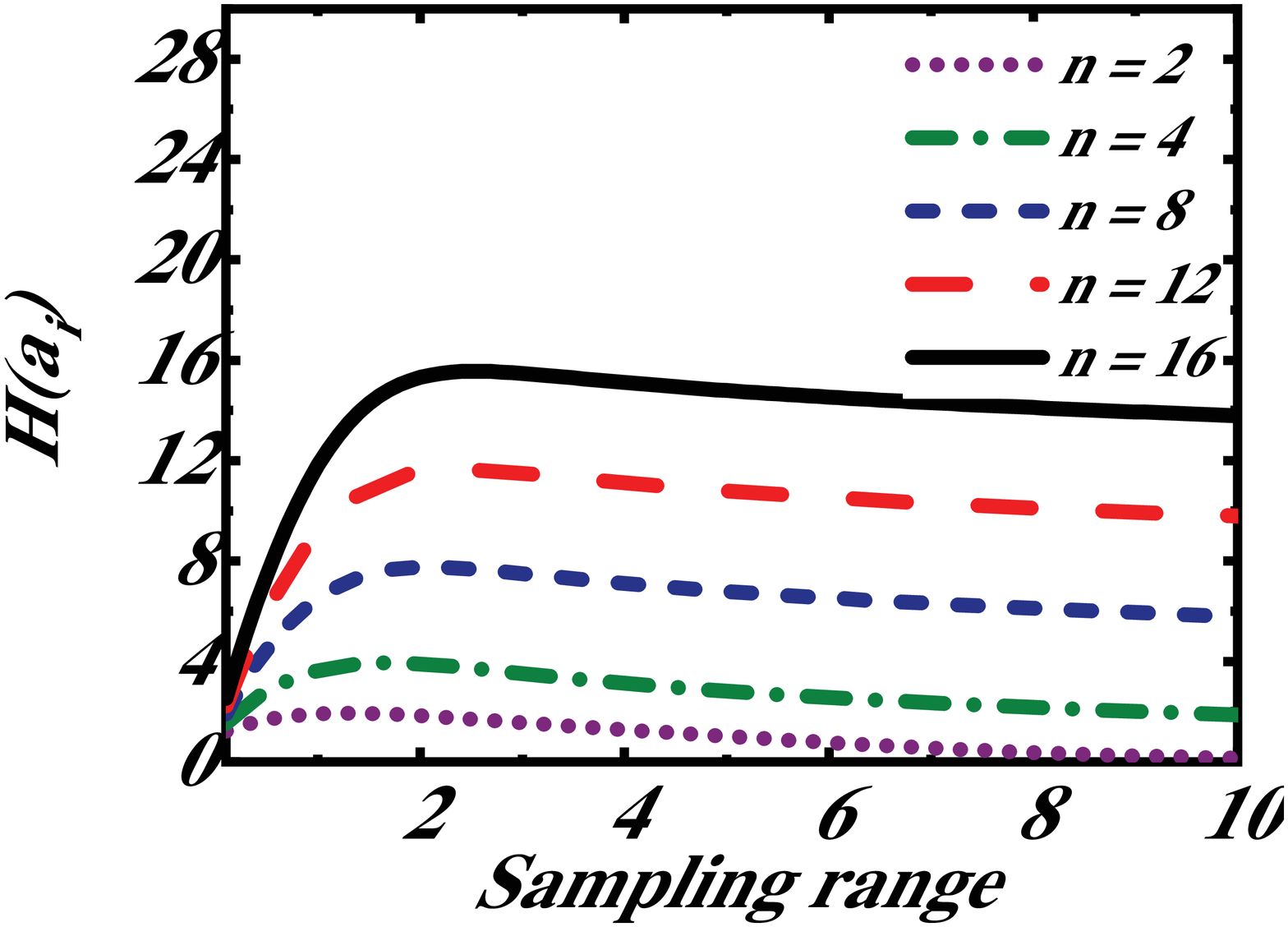}}
\subfigure[ ]{ \label{fig:subfig:b} 
\includegraphics[width=0.3 \textwidth]{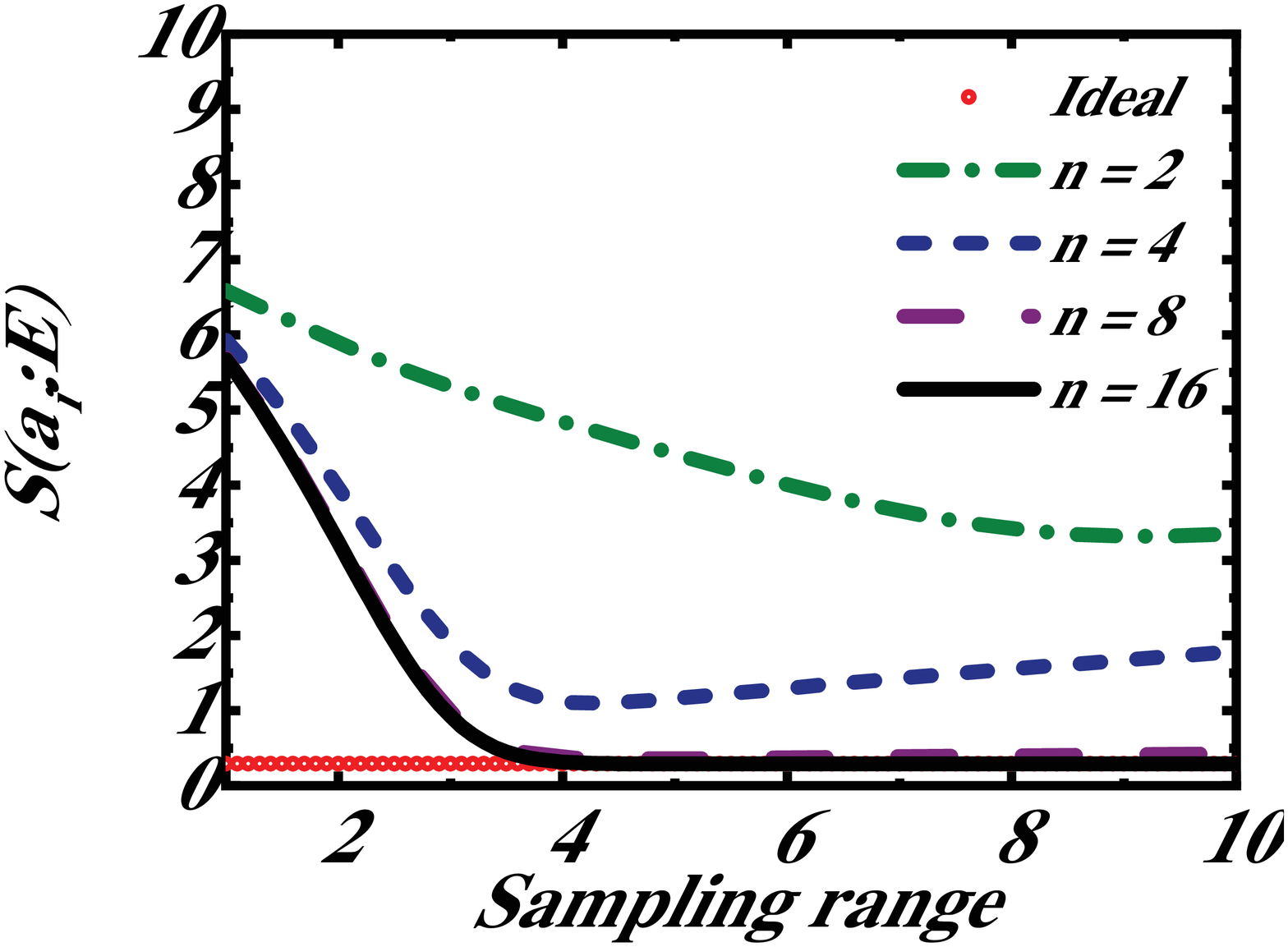}}
\subfigure[ ]{ \label{fig:subfig:c} 
\includegraphics[width= 0.3 \textwidth]{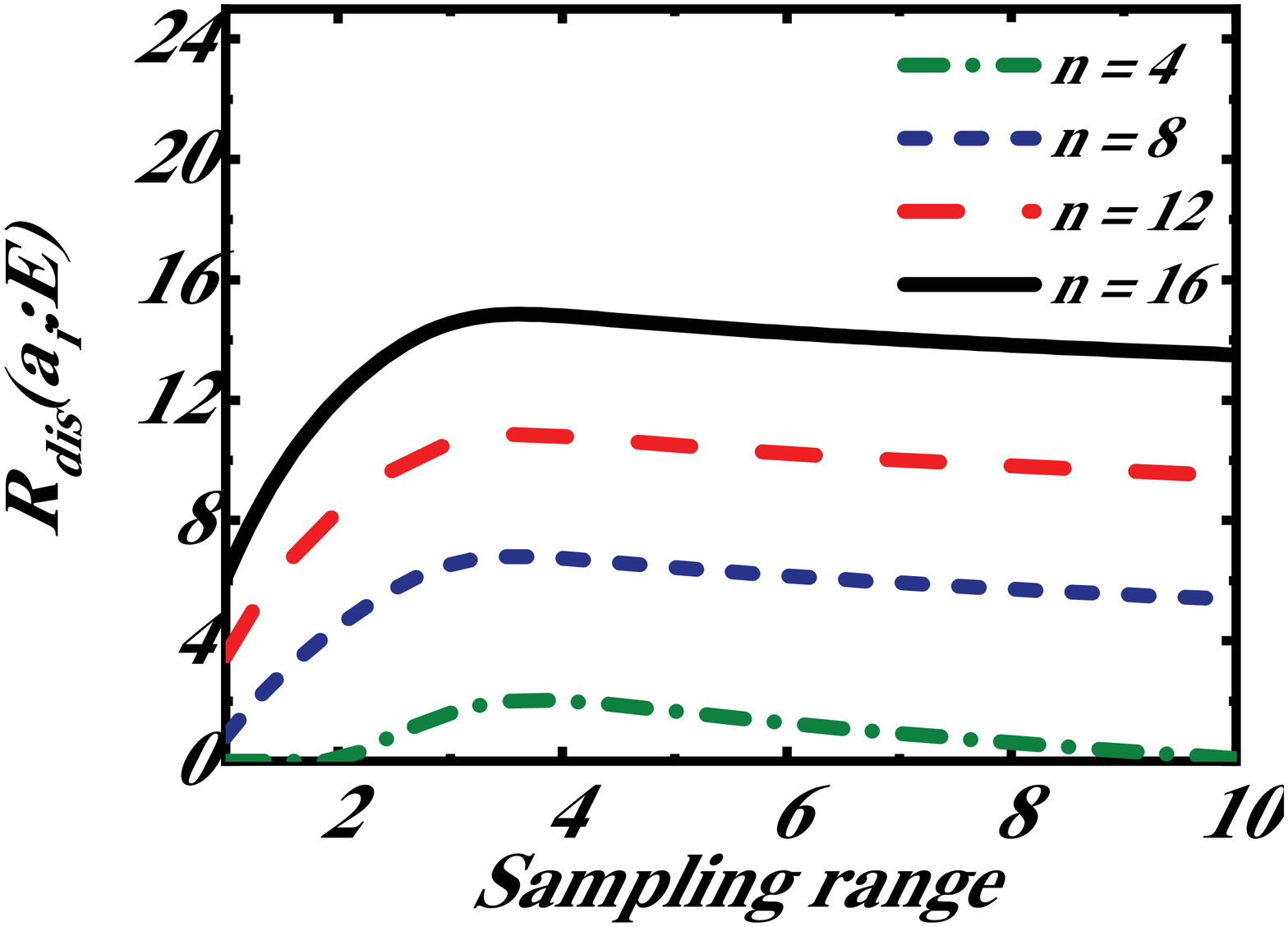}}
\caption{(color online) Simulation results for (a) Shannon entropy $H(a_i)$ of Alice's measurement results, (b) upper bound of Eve's information $S(a_i:E)$ and (c) extractable quantum randomness ${R_{dis}}({a_i}|E)$ as a function of sampling range $N$ with resolution $n=2,\ 4,\ 8,\ 12$ and $16$, respectively. The probability distribution of $a_i$  is simulated by Eq. (11) and $H({a_i}) =  - \sum\nolimits_{i = {i_{\min }}}^{i = {i_{\max }}} {{p_{dis}}({a_i}){\log_2}({p_{dis}}({a_i}))} $, based on which the upper bound of quadratures variance ${\overline V_x}$ (${\overline V_p}$) for  $\rho_A$ is estimated.  Then one has $S({a_i:E}) \le\frac{{\overline \lambda   + 1}}{2}{\log _2}\frac{{\overline \lambda   + 1}}{2} - \frac{{\overline \lambda   - 1}}{2}{\log _2}\frac{{\overline \lambda   - 1}}{2},~\overline \lambda   = \sqrt{ {\overline V_x} {\overline V_p} }$. The ideal case in (b) corresponds to the estimated upper bound of $S(a_i:E)$  when $N \to \infty ,n \to \infty$ as in Eq. (10). Finally, ${R_{dis}}({a_i}|E)$ is estimated by Eqs.~(8) and (9). The excess noise is chosen to be $\varepsilon  = 0.1$ and $\rho_A$ is assumed to be a symmetric Gaussian state with $V_x=V_p=1+\varepsilon$.}
\label{fig:Fig2} 
\end{figure}

The finite sampling range will make us lose the information about probability distribution of $a$ outside the detection range, directly influence the measured probability distribution $p_{dis}(a_i)$ given $p(a)$ (as in Appendix A), and thus affect the classical information $H({a_i})$ of measurement results (as in Fig.~\ref{fig:Fig2}(a)), the estimated upper bound ${\overline V_x}$ (as in Appendix C) and Eve's information $S(a_i:E)$ (as in Fig.~\ref{fig:Fig2}(b)).  For a fixed resolution $n$, the effects of dynamical sampling range $N$ are:
\begin{enumerate}[fullwidth]
\item If it is too small, the measurement outcomes become more predictable, and the Shannon entropy will reduce dramatically as in Fig.~\ref{fig:Fig2}(a) due to the oversaturated measurement results, which will compromise both the rate and the security of the random number generation. Meanwhile, one cannot estimate precisely the CM of the input states  $\rho_A$, i.e. overestimate the variance of $a$ based on $a_i$, thus overestimate Eve's information $S(a_i:E)$ (as in Fig.~\ref{fig:Fig2}(b)).
\item If it is chosen too large, most sampling bins will be unoccupied, and most measurement results lie in central bins, which will reduce the extractable randomness.
\end{enumerate}

Finally, the extractable randomness ${R_{dis}}({a_i}|E)$  increases significantly with $N$ when it is small, and reduces slowly after the optimal choice of $N$ as in~\ref{fig:Fig2}(c). In practice, the absolute value of $N$ for an ADC dvice is usually fixed. However, one can control the relative value of $N$ in shot-noise-unit (SNU) by adjusting the amplification parameter of homodyne detection (by controlling the power of LO or electronic amplifier). Usually, setting $N \in [3\sigma ,5\sigma ]$ in SNU will attain performance close to the optimal strategy.

\subsection{Effects of finite sampling resolution}

\begin{figure}[t]
\begin{center}
\includegraphics[width= 0.6 \textwidth]{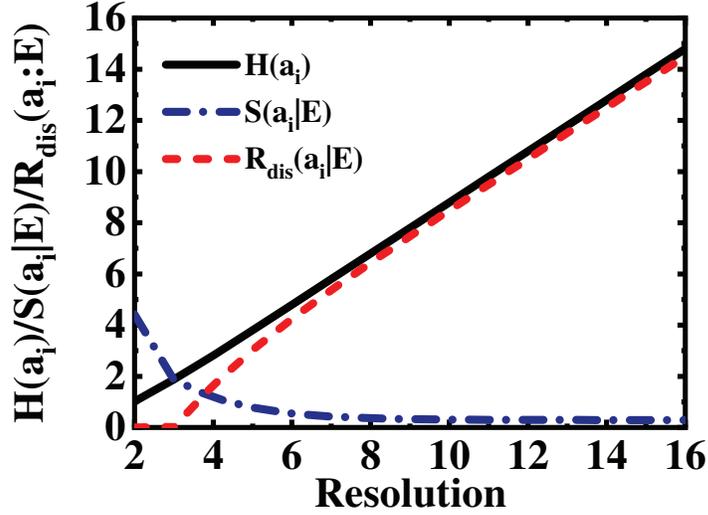}
\end{center}
\caption{(color online) Simulation results for Shannon entropy $H(a_i)$ of Alice's measurement results, upper bound of Eve's information $S(a_i:E)$ and extractable quantum randomness ${R_{dis}}({a_i}|E)$  as a function of resolution $n$ with sampling range $N=5$. The simulations of $H(a_i)$ and $S(a_i:E)$ follows the same methods as in Fig.~\ref{fig:Fig2}. The excess noise is chosen to be $\varepsilon  = 0.1$  and $\rho_A$ is assumed to be a symmetric Gaussian state with $V_x=V_p=1+\varepsilon$.}
\label{fig:Fig9-Rdis_vs__n}\label{fig:Fig3}
\end{figure}

The finite resolution $n$ will make us lose the information about probability distribution of $a$ inside discrete intervals $[(i-1/2)\Delta, (i+1/2)\Delta], i\in\{-2^{n-1},...,2^{n-1}-1\}$ (as in Appendix A). Given a fixed $N$, the larger the $n$, the more information can be got about $a$. It is clear that the performance of the protocol will increase with resolution $n$ as in Fig.~\ref{fig:Fig3}. Given a fixed sampling range $N$, the effects of resolution $n$ are:
\begin{enumerate}[fullwidth]
  \item If $n$ is small ($\Delta$ is large), one cannot estimate precisely the CM of $\rho_A$ (see Appendix C), thus overestimate the variance of quadratures for  $\rho_A$ and Eve's information $S(a_i:E)$. Furthermore, most measurement results lie in central bins (see Appendix A), which will reduce the Shannon entropy significantly (almost linearly) as in Fig.~\ref{fig:Fig3}.
  \item The classical information  $H(a_i)$ increases almost lineally with $n$, while the estimated Eve's information reduces with $n$. As a result, the total extractable randomness will increase with $n$ (as in Fig.~\ref{fig:Fig3}).
\end{enumerate}

In practice, given a fixed $N$, one should choose a larger $n$ to attain better performance.

\subsection{ Effects of excess noise}

\begin{figure}[htbp]
\centering
\subfigure[ ]{ \label{fig:subfig:a} 
\includegraphics[width=0.45 \textwidth]{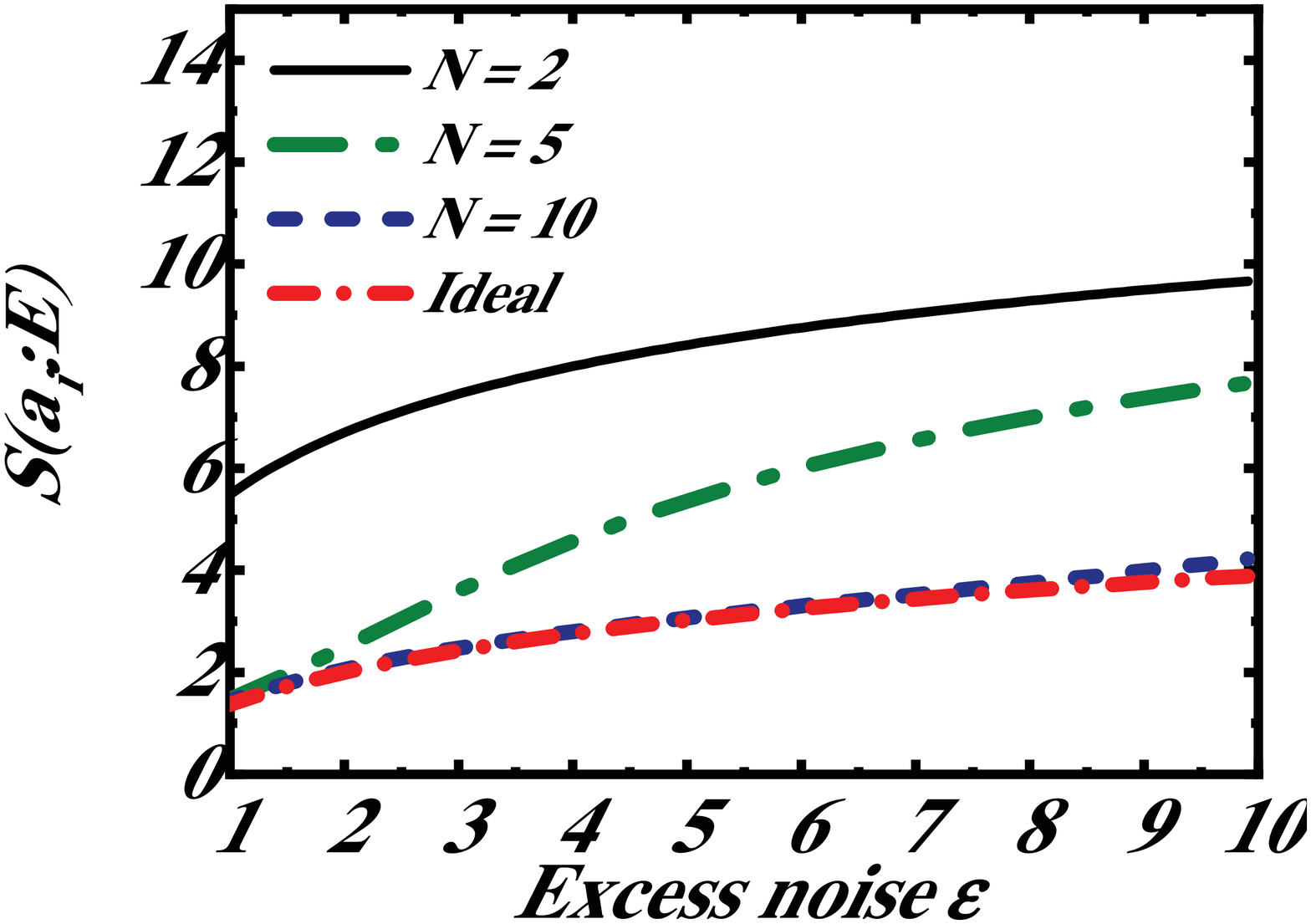}}
\subfigure[ ]{ \label{fig:subfig:b} 
\includegraphics[width=0.45 \textwidth]{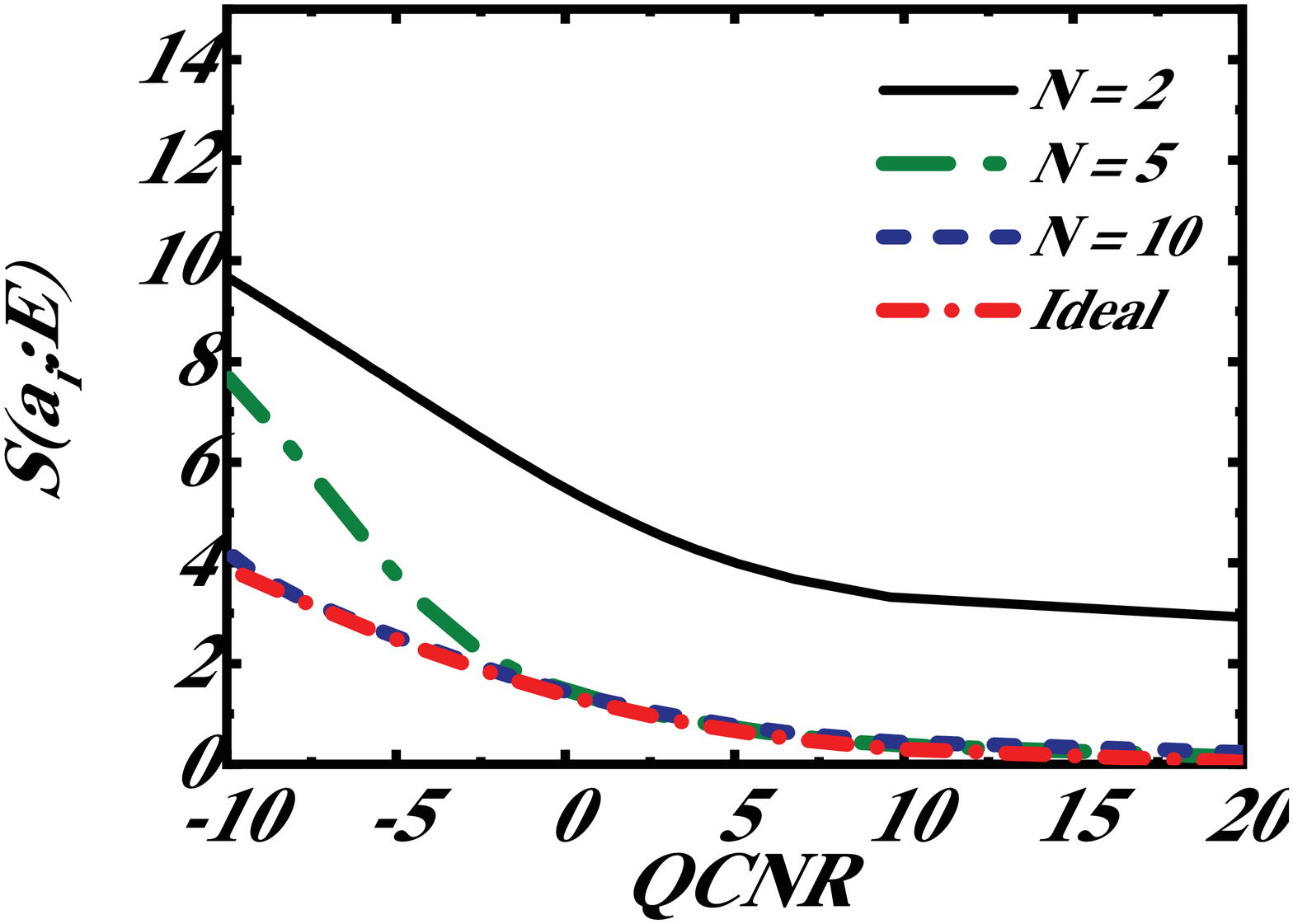}}
\caption{(color online) Simulation results for upper bound of Eve's information  $S(a_i:E)$ as a function of (a) excess noise and (b) QCNR, with $n=8$ and $N=2, 5, 10$, respectively. The ideal case corresponds to the estimated upper bound of $S(a_i:E)$  when $N \to \infty ,n \to \infty$. $\rho_A$ is assumed to be a symmetric Gaussian state with $V_x=V_p=1+\varepsilon$.}
\label{fig:Fig4-VonNeumannEntropy-vs_epsilion_QCNR_R} 
\end{figure}

\begin{figure}[b]
\centering
\subfigure[ ]{ \label{fig:subfig:a} 
\includegraphics[width=0.45 \textwidth]{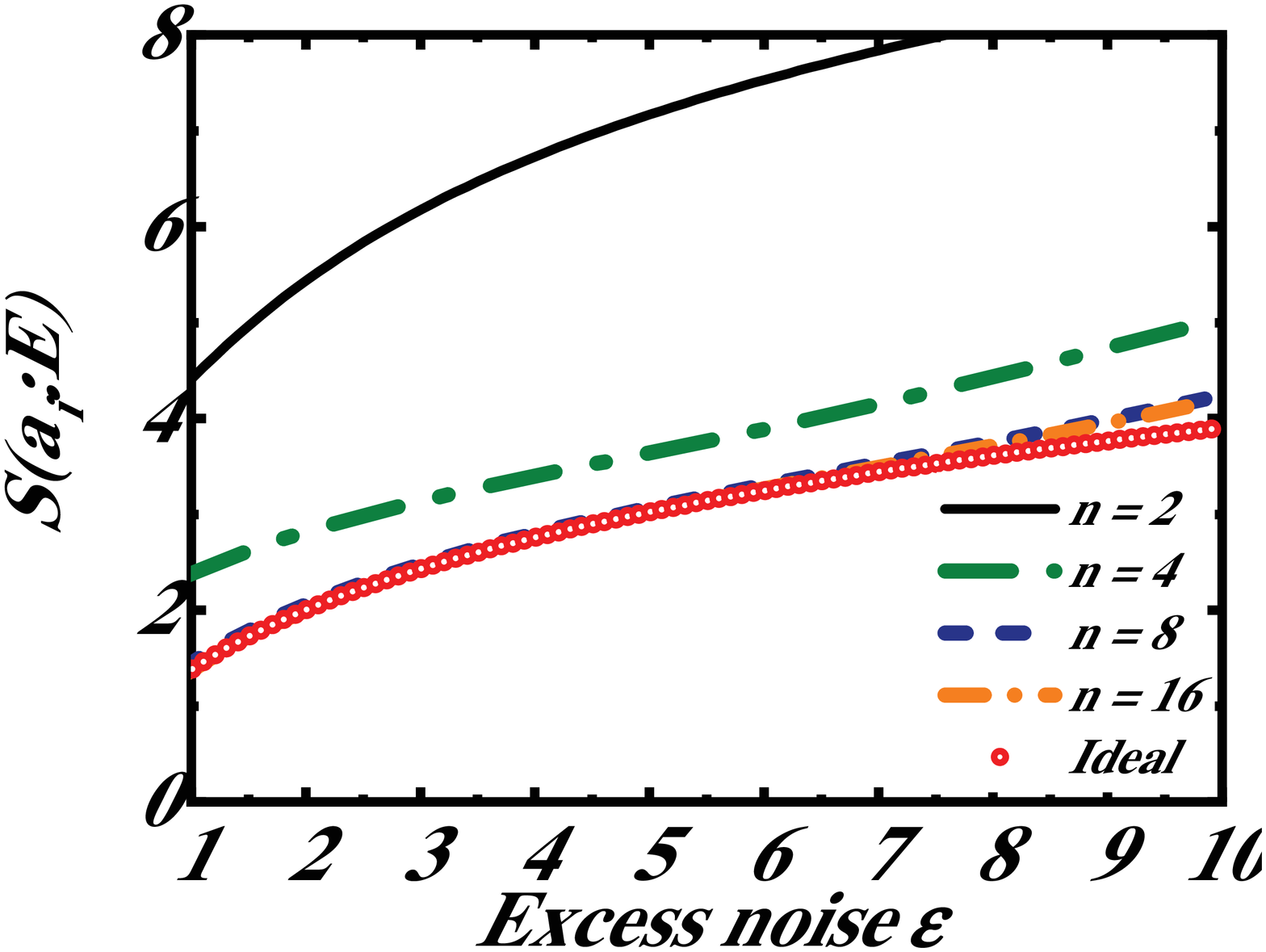}}
\subfigure[ ]{ \label{fig:subfig:b} 
\includegraphics[width=0.45 \textwidth]{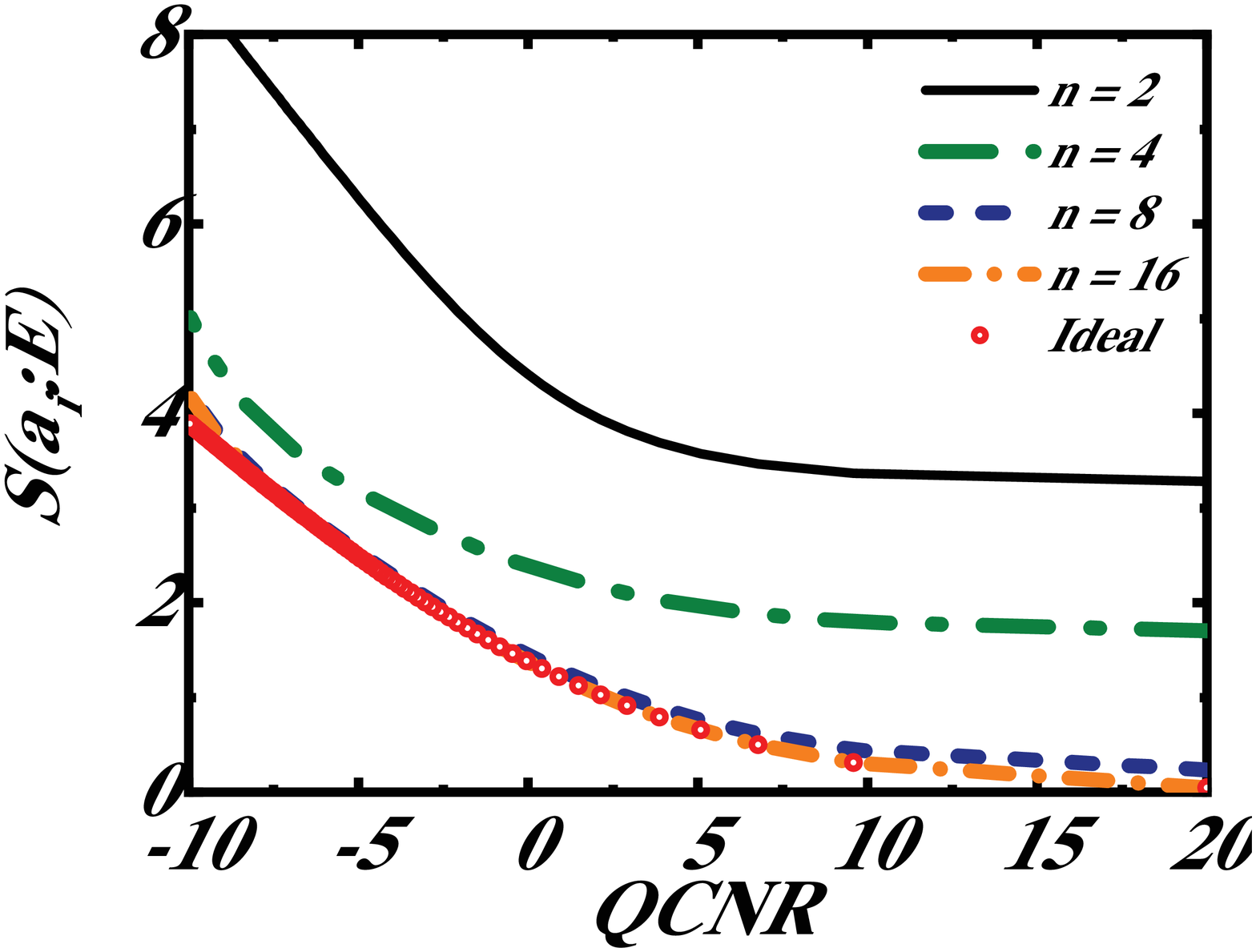}}
\caption{(color online) Simulation results for upper bound of Eve's information $S(a_i:E)$ as a function of (a) excess noise and (b) QCNR, with $N=10$ and $n=2, 4, 8, 16$, respectively.  The ideal case corresponds to the estimated value of $S(a_i:E)$  when $N \to \infty ,n \to \infty$. $\rho_A$ is assumed to be a symmetric Gaussian state with $V_x=V_p=1+\varepsilon$.}
\label{fig:Fig5-VonNeumannEntropy_vs_epsilion_QCNR_n} 
\end{figure}

\begin{figure}[t]
\begin{center}
\includegraphics[width= 0.6 \textwidth]{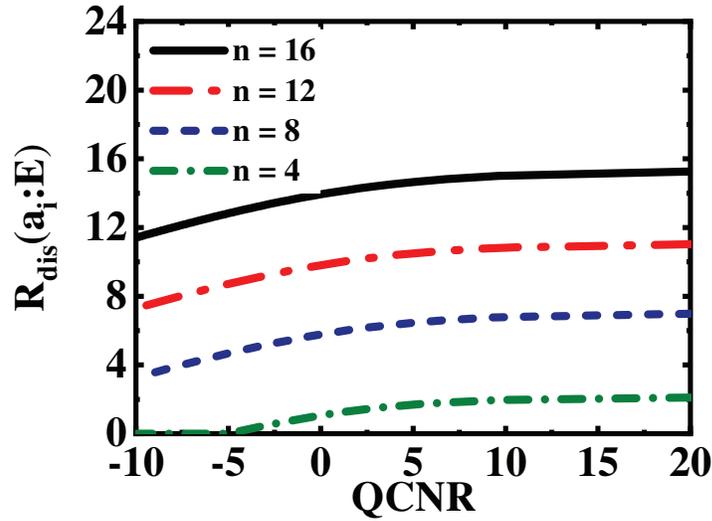}
\end{center}
\caption{(color online) Simulation results for extractable quantum randomness ${R_{dis}}({a_i}|E)$ as a function of QCNR with $n=4, 8, 12, 16$, respectively. The sampling range is chosen numerically optimally to be $N=3.3\sigma$. $\rho_A$ is assumed to be a symmetric Gaussian state with $V_x=V_p=1+\varepsilon$.}
\label{fig:Fig6-Rdis-vs_QCNR_n}
\end{figure}

\begin{figure}[b]
\centering
\subfigure[ ]{ \label{fig:subfig:a} 
\includegraphics[width=0.45 \textwidth]{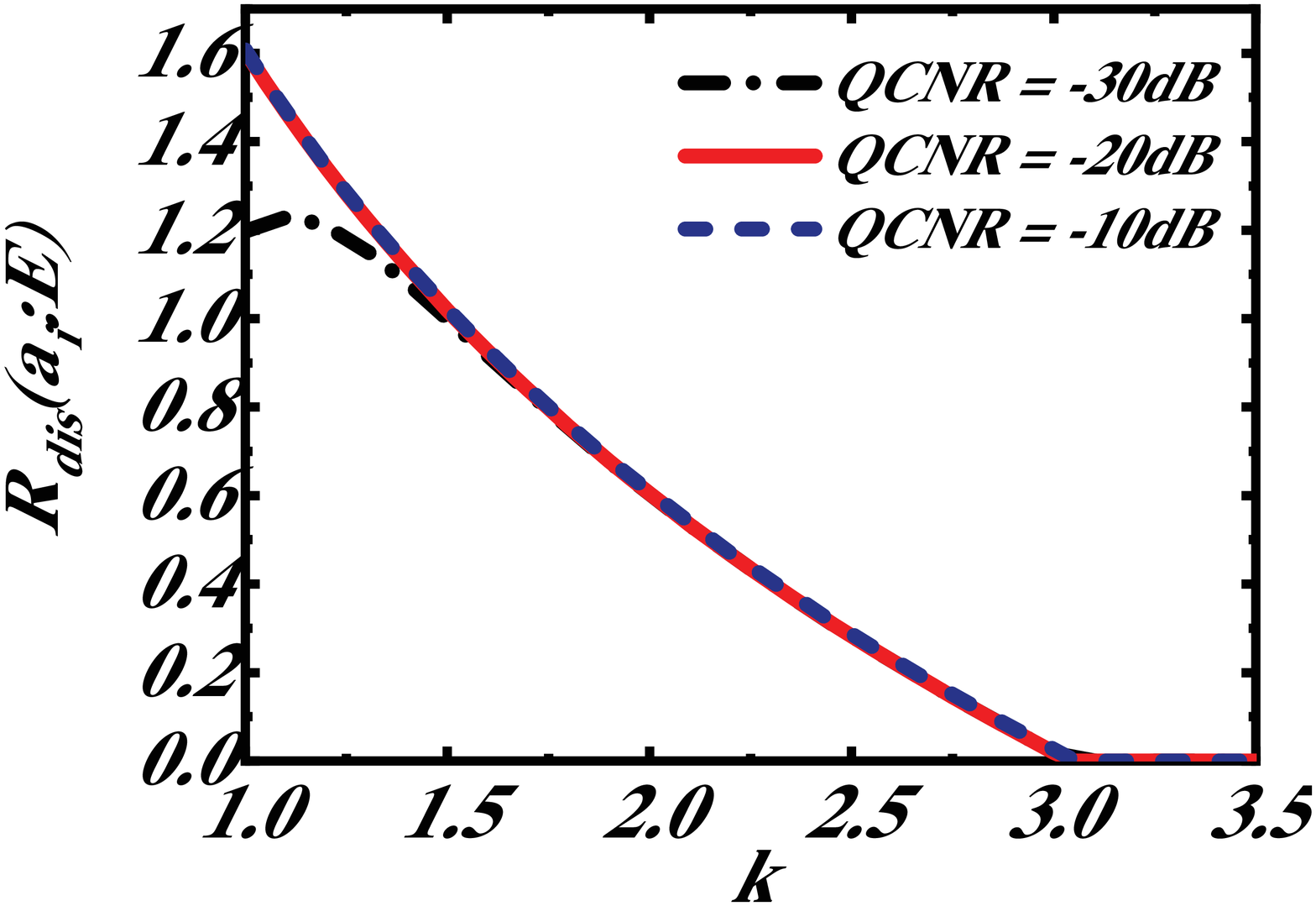}}
\subfigure[ ]{ \label{fig:subfig:b} 
\includegraphics[width=0.45 \textwidth]{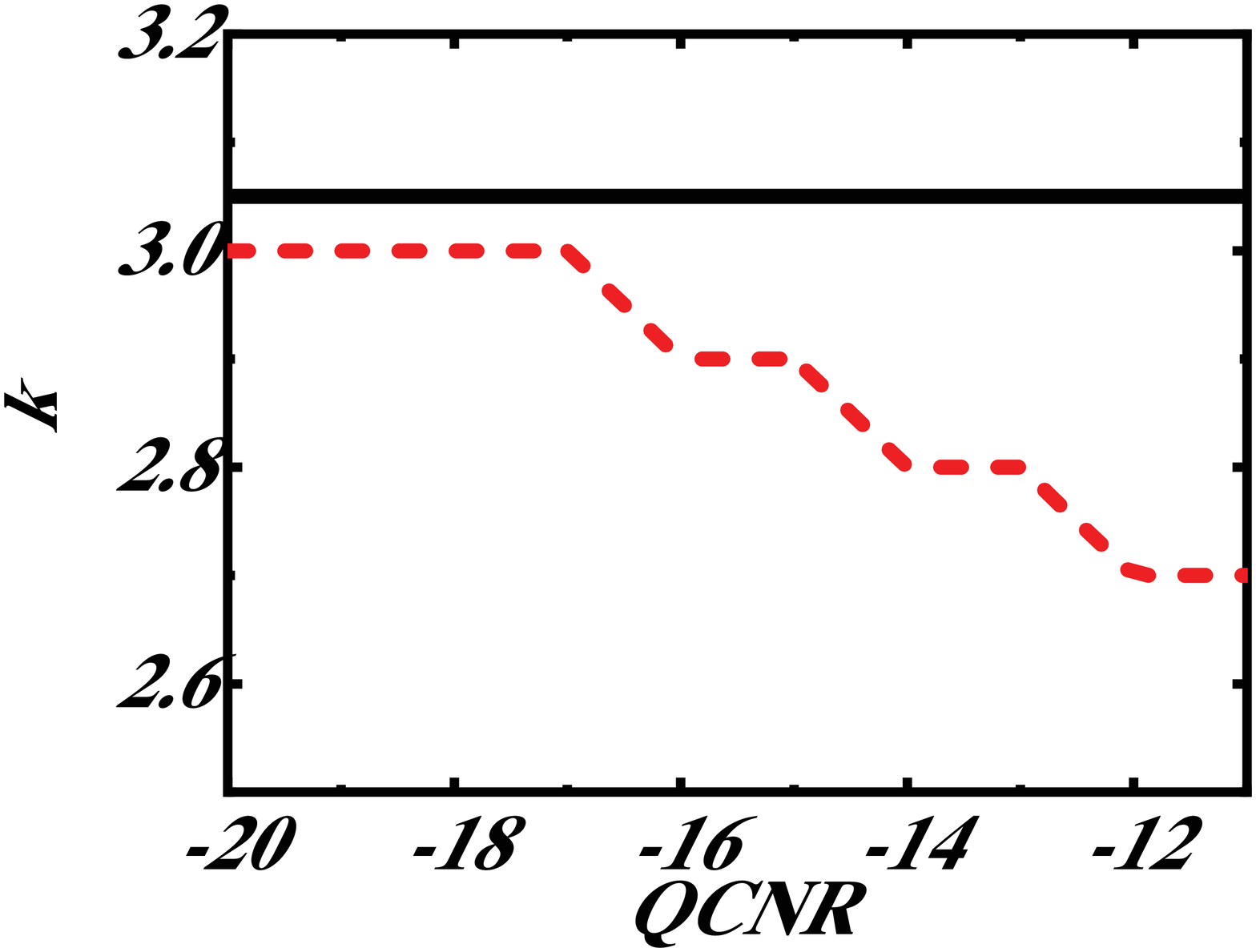}}
\caption{(color online) (a) Simulation results for extractable quantum randomness ${R_{dis}}({a_i}|E)$ as a function of $k=\Delta\sigma$ with $n=16$ and QCNR$=-10 dB,\ -20 dB,\ -30dB$, respectively. Quiet similar results are found for $n=8,\ 12$. In the simulations, we ignore the effects of digitization on estimation of CM to give an insight into the effect of parameter $k$. (b) The maximal allowed resolution $\Delta_{\max}=k/\sigma$ for low QCNRs $(\le -10dB)$ with $n=16$. Quiet similar results are found for $n=8,\ 12$. The effects of digitization on estimation of CM are ignored in black solid line and considered in red dashed line.}
\label{fig:Zeropoint}
\end{figure}
In our protocol, the detection is assumed to be trusted and well characterized, and all the excess noise is due to a quantum correlated Eve. The excess noise or QCNR mainly decides the correlation between Eve's quantum state and Alice's measurement results, which is a key parameter in security analysis. It is clear that $S(a_i:E)$  increases (decreases) with $\varepsilon$ (QCNR), and one needs to choose proper $N$ and $n$ to get a tighter upper bound on $S(a_i:E)$. For a given $n$, the optimal value $N$ varies with different QCNR, as in Fig.~\ref{fig:Fig4-VonNeumannEntropy-vs_epsilion_QCNR_R}. For given $N$ and $\varepsilon$, one can obtain a tighter bound of $S(a_i:E)$ with larger $n$, as in Fig.~\ref{fig:Fig5-VonNeumannEntropy_vs_epsilion_QCNR_n}. Even when excess noise is much larger than quantum noise, one still can get a tight bound on Eve's information.

The final performance of the protocol is resistant against to the excess noise as shown in Fig.~\ref{fig:Fig6-Rdis-vs_QCNR_n}. More surprisingly, even if the QCNR goes below $0$ (e.g. -10 dB), that is, excess noise due to Eve becomes larger than quantum noise, one can still obtain a nonzero number of certified random bits that are independent of Eve's side information. This means one can use high bandwidth commercial balanced receivers which does not require the receiver's high QCNR (e.g. $\ge 10dB$) as in former CV-QRNG experiment based on vacuum fluctuation~\cite{QRNG_Vac1,QRNG_Vac2,QRNG_Vac3} or CV-QKD experiment, and thus dramatically increases the random number generation rates.

An interesting and important question is what is the relationship between the system configuration and the maximal tolerable excess noise, or equivalently, the lowest QCNR that still keeps the extractable randomness non-negative. Intuitively, no matter how low the QCNR is, there is always some quantum randomness existing in the measurement results, which can be extracted as long as the resolution $n$ is high enough, \emph{i.e.}, at least to let the quantum randomness part change one bit of the measurement result. Considering that the Gaussian state extremality theorem and the assumption that Eve holds the purification of the whole state are used to estimate $R_{dis}\left(a_i|E\right)$, it is natural to `\emph{imagine}' the whole state is a two-mode squeezed state with variance $\sigma^2$. In principle, Eve could provide $X$-squeezed states to Alice to gain more advantage than vacuum states, due to its reduced variance $V^{sq}_x = 1/{\sigma^2}$, since the final random bits are extracted only from the $X$-quadrature measurements. Thus, we predict that, if the quantum randomness are contained in two or more bins, \emph{i.e.}, $\left[-\Delta_{max},+\Delta_{max}\right]\in\left[-3\sqrt{V^{sq}_x},+3\sqrt{V^{sq}_x}\right]~\left( \to \Delta_{max} \le \frac{k}{{\sigma }},k\approx 3 \right)$,
then $R_{dis}\left(a_i|E\right)$ will be non-negative.

We check the prediction numerically with conditions $n=8,12,16$ and different QCNRs. In Fig. \ref{fig:Zeropoint} (a), $R_{dis}\left(a_i|E\right)$ with several typical QCNRs ($-10dB,-20dB,-30dB$) are shown, and it is found $k\approx 3.05$, which is almost the same for other different QCNRs (see the black solid line in Fig. \ref{fig:Zeropoint} (b)). However, if taking into consideration the modification of ${\overline V_x}$ as in Appendix C, it requires higher resolution $n$ (smaller bin width $\Delta_{max}$, see the red dashed line in Fig.~\ref{fig:Zeropoint} (b)).

\subsection{Effects of squeeze factor}

\begin{figure}[b]
\begin{center}
\includegraphics[width=0.6\textwidth]{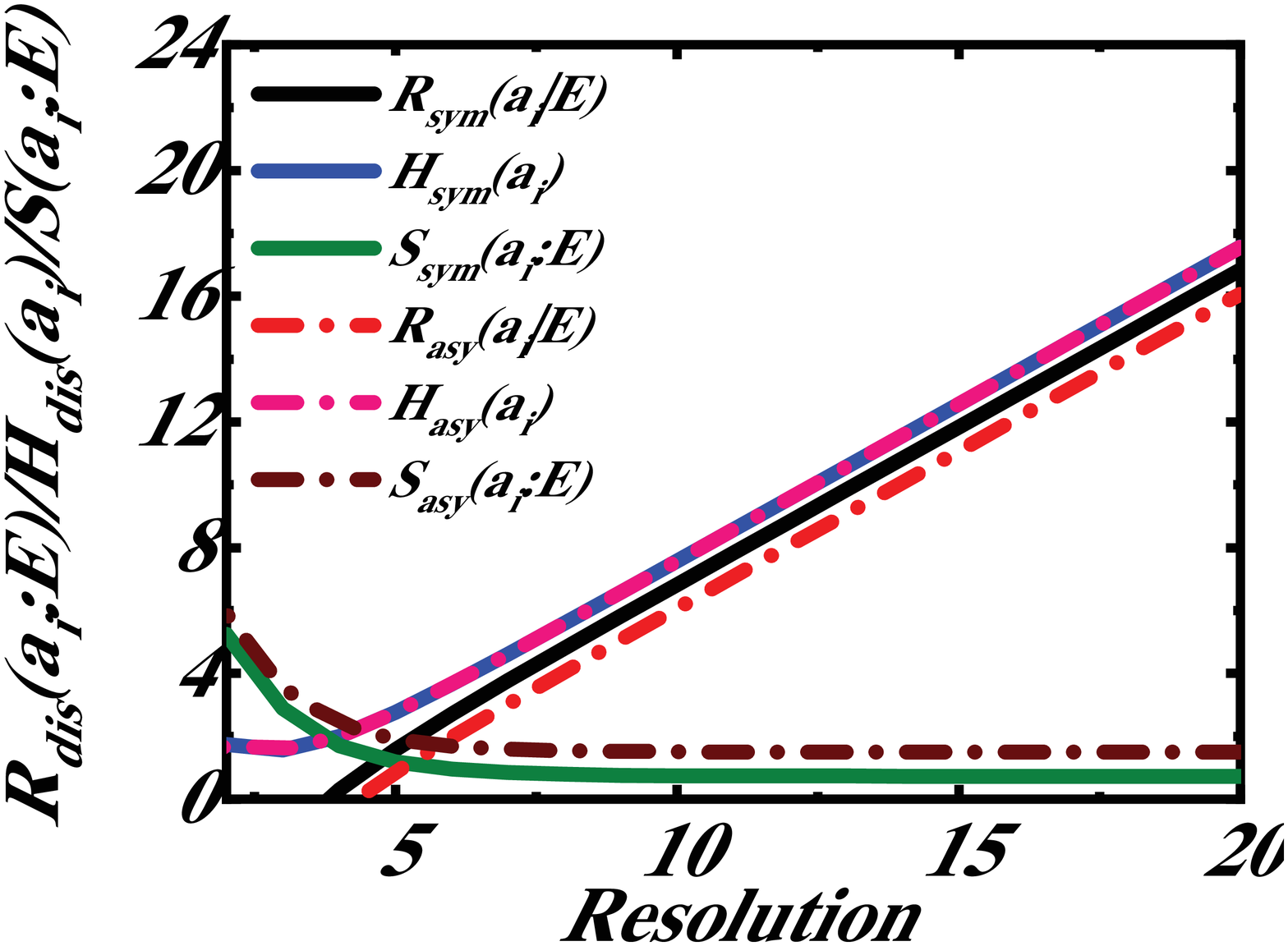}
\end{center}
\caption{(color online) Simulation results for extractable quantum randomness ${R_{dis}}({a_i}|E)$, Shannon entropy $H_{dis}(a_i)$ of Alice's $P$-quadrature measurement results and upper bound of Eve's information $S(a_i:E)$ as functions of resolution $n$. The solid lines are the case of symmetric Gaussian state (vacuum state) and the dashed lines are the case of asymmetric Gaussian state (squeezed state with squeeze factor of 13$dB$). The excess noise is chosen to be $\varepsilon  = 0.1$.}
\label{squeezed_state_case}
\end{figure}

In practice, the state prepared by Eve could not be a vacuum state with some excess noises, but may be asymmetric between two quadratures $x$ and $p$, which refers to the squeezed-state case. We denote the squeeze factor by $r$ as an example and the CM of ${\rho _A}$ can be modified by
\begin{equation}
{\gamma _A} = \left[ {\begin{array}{*{20}{c}}
{{e^{ - 2r}} + \varepsilon }&0\\
0&{{e^{2r}} + \varepsilon }
\end{array}} \right],
\end{equation}
where we assume that $x$ is the squeezed quadrature with variance ${V_x} = {e^{ - 2r}} + \varepsilon$ and $p$ is the anti-squeezed quadrature with variance ${V_p} = {e^{2r}} + \varepsilon$.

For simplification, we assume the anti-squeezed quadrature $p$ is exploited to generate raw key and both quadratures $x$ and $p$ are used for entropy estimation. In other words, the discrete shannon entropy $H_{dis} \left( {{a_i}} \right)$ can be estimated by the data only from $P$-quadrature measurements, and meanwhile both data from $X$-quadrature measurements and $P$-quadrature measurements are needed to estimate the upper bound $S\left( {\rho _A^G} \right)$. In order to estimate the upper bound of two variances ${\overline V_x}$ and $ {\overline V_p}$ as accurately as possible, and thus $\overline \lambda  = \sqrt {{\overline V_x} {\overline V_p}}$ can be calculated precisely, we set different sampling ranges ${N_x} = 3{\sigma _x}$ and ${N_p} = 3{\sigma _p}$ for two quadrature measurements to obtaining almost optimal performance.

We assume that Eve prepares squeezed states with practical feasible parameters as an example, where the variances are ${V_x} = 0.05 + \varepsilon$ and ${V_p} = 20 + \varepsilon$, referring to squeeze factor of 13$dB$ \cite{Opt.Lett.43.110.2018}. The performance of the asymmetric Gaussian state (squeezed state) is shown in Fig. \ref{squeezed_state_case}, and it is compared to the protocol using symmetric Gaussian state (vacuum state). The estimation method of the asymmetric-state case is the same as the symmetric-state case except for a small difference in the estimation of the upper bound of Eve's information $S(a_i:E)$, where two variances used in estimating $\overline \lambda$ are different.

Simulation results indicate that, the Shannon entropy $H_{dis}(a_i)$ of Alice's measurement results are the same under the same sampling bits, considering both symmetric-state case and asymmetric-state case. However, because Eve's information $S(a_i:E)$ under asymmetric-state case is larger than that of symmetric-state case, the extractable randomness ${R_{dis}}({a_i}|E)$ of asymmetric-state case (red solid line) is a little smaller than that of symmetric-state case (black dashed line). This result can be obtained directly by simplifying $\lambda$, which reads
\begin{equation}
{\lambda ^2} = {V_x}{V_p} = 1 + \varepsilon \left( {{e^{ - 2r}} + {e^{2r}}} \right) + {\varepsilon ^2},
\end{equation}
and $\lambda$ takes the minimum value when $r$ is 0, which refers to the case of symmetric states without squeezing operation. $S(a_i:E)$ is a monotonic incremental function about $\lambda$, which means that Eve's amount of information is only minimal without squeezing and any asymmetry of the quantum states ${\rho _A}$ will decrease the final extractable randomness. The result demonstrates that the QCNR of the squeezed quadrature is much smaller than that of non-squeezed case, which will inevitable affect the estimation of the CM.

\section{Off-line and Real-time Experimental Implementations and Performances}
\begin{figure}[t]
\begin{center}
\includegraphics[width=0.6\textwidth]{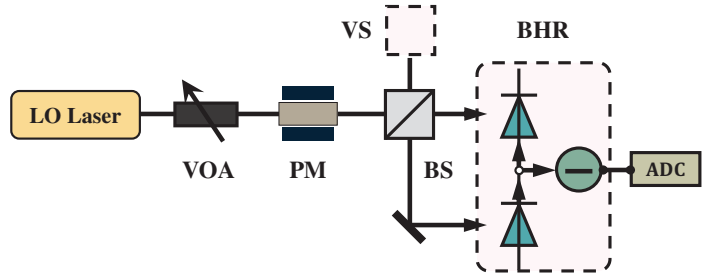}
\end{center}
\caption{(color online) Experimental setup of the proposed CV-SI-QRNG. LO: local oscillator, VOA: variable optical attenuator, PM: phase modular, BS: 50/50 beam splitter, VS: vacuum state, BHR: balanced homodyne receiver, ADC: analog-to-digital converter. The CV-QRNG is realized with off-the-shelves components.}
\label{fig:Fig8-Experimentalsetup}
\end{figure}

To validate the proposed protocol, a CV-SI-QRNG experimental setup is built based on balanced homodyne receiver to measure the vacuum fluctuations as in Fig.~\ref{fig:Fig8-Experimentalsetup}. We implemented an all-in-fiber setup with off-the-shelves devices. The local oscillator (LO) is a narrow line 1550 nm laser (Thorlabs SLF1550P, linewidth $50$ KHz), and the LO power is carefully adjusted to obtain the optimal performance. The 50:50 beamsplitter (BS) brings LO signal interfered with the vacuum states to the balanced receiver (Thorlabs PDB480C, bandwidth 1.6GHz). The phase of LO is randomly shifted between $0$ and $\pi/2$ by a phase modulator (PM) to realized the random sampling between $X$ and $P$ quadratures measurements, based on an initial random seed $t$. Finally, the measurement results of the balanced receiver are sampled in real-time by a 12-bit ADC (TI ADC12D1800, bandwidth 3.5GHz) with a sampling rate of 1.8G SPS to acquire the raw data, which is to be analyzed by the proposed model to extract secure randomness.

\begin{figure}[b]
\begin{center}
\includegraphics[width=0.6\textwidth]{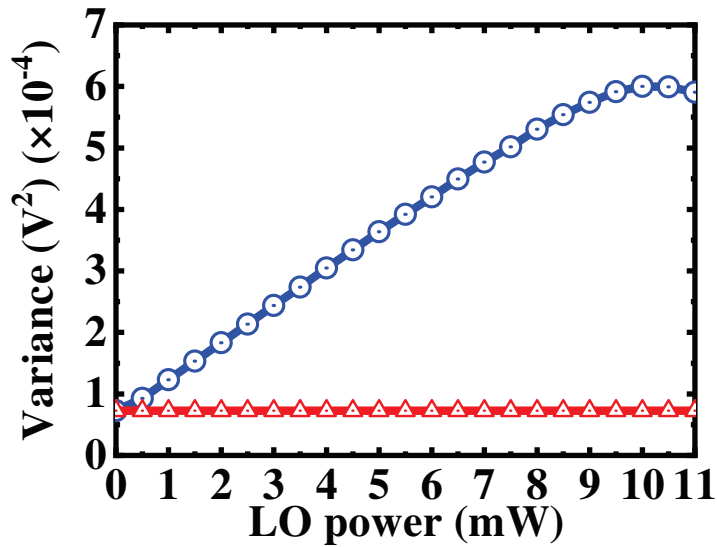}
\end{center}
\caption{(color online) Variance vs LO power. This figure shows the voltage variance of the sampled raw data as a function of the LO power. In the region from 0mW to 9.5 mW, it shows a relatively clear linearity between the voltage variance and the sampled raw data. While the LO power increases higher than 9.5 mW, the detection and amplification in the balanced receiver starts to saturate, resulting in the decrease of the linearity, and the peak value of the voltage variance is obtained at 10 mW.}
\label{fig:Fig9-LO_power_VS_Variances}
\end{figure}

To obtain optimal performance, the LO power is increased from 0 mW with a fixed step of 0.5 mW and the voltage variance of the raw data corresponds to each LO power value is calculated and recorded, which is shown in Fig.~\ref{fig:Fig9-LO_power_VS_Variances}. In the region from 0 mW to 9.5 mW, the variance of the sampled raw data enhances linearly with the increase of the LO power and the peak value is observed at the LO power value of 10 mW due to the saturation of the balanced homodyne receiver. The LO power is fixed at 10 mW in the experiment to obtain optimal performance. The variance of the sampled raw data has a non-zero value even when the LO power is turned off, which is generally attributed to the classical noise resulted from the electromagnetic disturbance, the thermal fluctuations, the imperfection of the experimental setup and even the manipulation of eavesdroppers. In practical implementation, the classical noise can hardly be eliminated and will also be sampled into the raw data, resulting in an equivalently impure quantum states and impairing the randomness and security.

\begin{figure}[t]
\begin{center}
\includegraphics[width=0.6\textwidth]{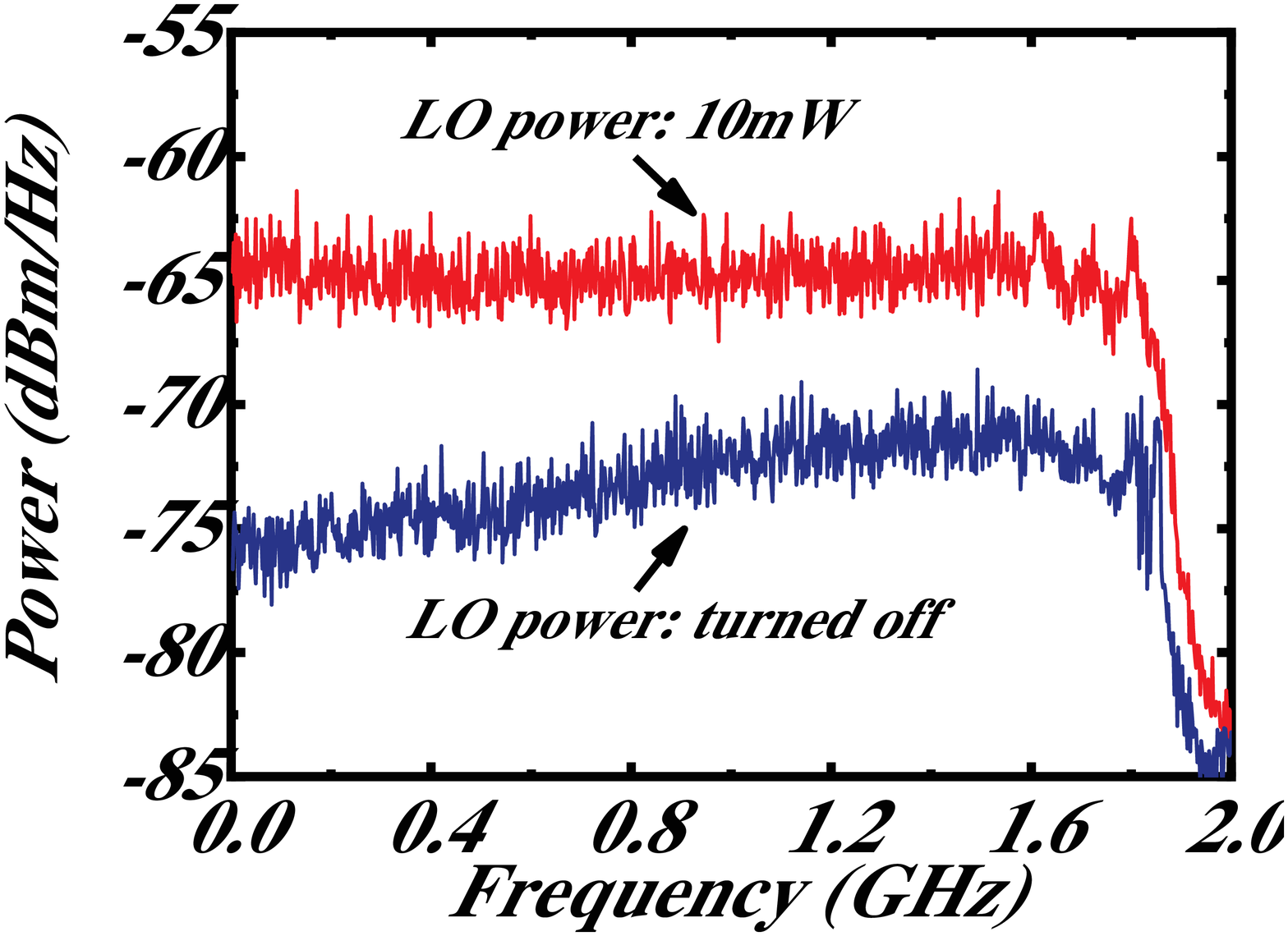}
\end{center}
\caption{(color online)  The power spectal density of the vacuum fluctuations when the LO power is 10mW (red line) and the electrical noise when the LO power is turned off (blue line). Within the detected frequency range, the vacuum fluctuations dominate in terms of the power.}
\label{fig:Fig10-FD}
\end{figure}

\begin{figure}[b]
\begin{center}
\includegraphics[width=0.6\textwidth]{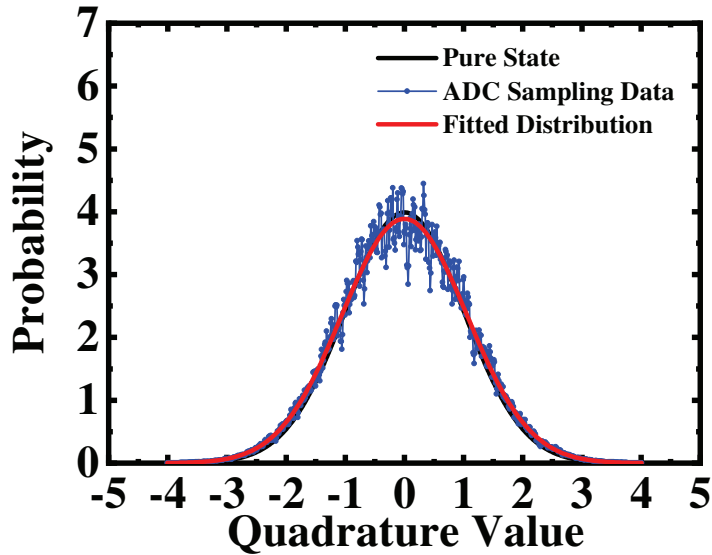}
\end{center}
\caption{(color online) The probability distribution of the ideal pure vacuum state (black line), the probability distribution of the ADC sampled raw data (blue line) in shot-noise-unites and the fitted distribution for raw data (red line).}
\label{fig:Fig11-Fitline}
\end{figure}

To measure the vacuum fluctuations with respect to the excess noise, the measurements in the frequency domain have been performed by using an RF spectrum analyzer, which is shown in Fig.~\ref{fig:Fig10-FD}. Two different spectra have been acquired: the vacuum fluctuations when the LO power is 10 mW (red line) and the electrical noise when the LO power is turned off (blue line). From the figure we can see that in the detected range, the power of vacuum fluctuations is obviously higher than that of the electrical noise (the average gap between them is about 8.37dB within 0$\sim$1.6 GHz), which means the vacuum fluctuations dominate the output and demonstrates the effectiveness of the detection.

We present the results obtained on a typical run of  ${n_{tot}} = 2.6214 \times {10^9}$ data samples. The corresponding measured shot noise variance is $5.5572\times 10^{-4}\ V^2$, excess noise variance is $6.31\times 10^{-5}\ V^2$, and the total measurement results variance of $X$ quadratures is $6.1182\times 10^{-4}\ V^2$. In SNU, the variance of the pure vacuum state is normalized to 1, while the experimentally measured $X$ quadrature variance is $\sigma^2=1.1135$ as shown in Fig.~\ref{fig:Fig11-Fitline}. As mentioned above, the deviation is mainly resulted from the classical noise in our experiment, which cannot be separated from the vacuum fluctuations in measurement results, leading to the impurity of the sampled raw data and the potential impairment of the security. The sampling range $N=21.2098=20.0998\sigma$ and excess noise $\varepsilon=0.1135$ in SNU, and the resolution $n=12$. The estimated upper bound ${\overline V_x}  = 1.1218$ and ${\overline V_p}= 1.1220$ in real experiment with the methods in Appendix C (the corresponding theoretical simulation results based on our model is ${\overline V_x}={\overline V_p}=1.1223$ for both quadratures under Gaussian assumption as in Eq. (11)).  The experimental estimation and corresponding theoretical simulation of Shannon entropy, von Neumann entropy and extractable randomness are $H({a_i})=8.7117(Exp)/8.7180(Theo)$, $S(\rho _A^G) = 0.3366(Exp)/0.3373(Theo)$ and ${R_{dis}}({a_i}|E) = 8.3751(Exp)/8.3807(Theo)$, respectively. Furthermore, the number of bits necessary for the switching between the two quadratures must be accounted. Following~\cite{QRNG_SDI2}, we set ${n_c}$$ = $$\sqrt {{n_{tot}}}$. Out of the $n_{tot}$ measurements, the check instants can be chosen in $\left( {\begin{array}{*{20}{c}}
{{n_{tot}}}\\
{{n_c}}
\end{array}} \right)$ different ways, which can be encoded in a seed $t = \left\lceil {{{\log }_2}\frac{{{n_{tot}}!}}{{{n_c}!({n_{tot}} - {n_c})!}}} \right\rceil$ bits long. The final secure generation rate, i.e. true random bits per measurement, is
\begin{equation}
{R_{\sec }} = \frac{1}{{{n_{tot}}}}\left[ {({n_{tot}} - {n_c})(H({a_i}) - S({a_i}:E)) - t} \right].
\label{key_rate_asymptotic}
\end{equation}
Given ${n_{tot}} = 2.6214 \times {10^9}$, we employed ${n_c} = 5.12 \times {10^4}$ bits to evaluate the extractable randomness, and $t = 8.7482 \times {10^5}$. In our experiments, the corresponding secure rate of every measurement is $R_{sec} = 8.3746$ bits with sampling rate 1.8G SPS, which indicates an equivalent secure bit generation rate of $15.07$ Gbits/s. The final generated random bits sequences have passed all the NIST and DIEHARD tests. It should note that, Eq. (\ref{key_rate_asymptotic}) strictly holds under the asymptotic-limit case, where the total block size $n_{tot}$ tends to infinity. Therefore our randomness generation rate is, rigorously speaking, a asymptotic rate without considering the finite-size effect, which will lead to a bias in estimating the CM ${\gamma _A}$, resulting in the estimation of $S({a_i}:E)$ deviating from the asymptotic-limit case. We leave the finite-size analysis of the randomness extraction for future investigations.

For the real-time implementation, we have also developed a parallel algorithm of Toeplitz hashing post-processing method on the field programmable gate array (FPGA) recently \cite{real-time}. The FPGA-based hardware can support our system to achieve a generation rate of $6$ Gbits/s under the existing hardware conditions, which is important for the fields requiring immediately available random numbers.

\section{Conclusion and Discussion}

We have proposed and experimentally demonstrated a CV-SI-QRNG protocol even if the source is untrusted or controlled by Eve. Based on the extremality of Gaussian states, a new theoretical model to estimate the lower bound on the extractable quantum randomness is established, which is similar to the security analysis of CV-QKD. The protocol is resistant to classical noise and losses (can be easily compensated by increasing LO power), which is beneficial for practical applications. The random numbers are sampled in real-time by a dedicated ADC hardware rather than oscilloscope~\cite{QRNG_SDI2}, which is beneficial  to a practical QRNG design. We experimentally demonstrate the protocol with commercial devices, and the final secure random number generation rates reach up to 15 Gbits/s in off-line and 6 Gbits/s in real-time respectively, and shows feasibility of the protocol with an ultra-fast, cheap and compact CV-QRNG. By using high bandwidth commercial balanced receivers and fast LO phase shifter, the secure generation rate can be increased to tens of Gbits/s.

The finite-size analysis should be investigated in further research, where one can use the fruitful theoretical tools built in CV-QKD~\cite{cvqkd_review}. The SI-QRNG protocol proposed here is actually a CV-QKD protocol with a trivial sender, who always sends the vacuum state. Therefore, one can follow the same universal composable framework (UCF) in Ref.~\cite{Finite_size1,Finite_size2,Finite_size3,Finite_size4}. However, the finite-size analyze of SI-QRNG has two main differences with CV-QKD under UCF, i.e., no need for error correction, and only one parameter is statistically counted and tested in parameter estimation test. The core of finite-size analysis is to evaluate Eve's information about the measured random bit sequence, represented by a quantity of smooth min-entropy $H^{\varepsilon'}_{\min}(a^{n_Q}_i|E)_{\rho^{n_Q}}$ that varies with data block size. Moreover, the comparison between our protocol and the protocol given in Ref.~\cite{QRNG_SDI2} is a very interesting and important topic, which needs to be further studied considering the practical issues, such as the finite-size effect, the finite sampling resolution and range of the detector.

Furthermore, in this paper, the LO power is assumed to be constant and fixed and security analysis against the LO power fluctuation needs to be further investigated. In our lab experiment, the LO power is relatively stable. However, in practical application, it could be influenced by the environment or even by Eve.

\emph{Note added}. Recently, an independent work has been published in Ref.~\cite{QRNG_SDI_HET}. This work also proposed a new analysis method for SI-QRNG, which also has experimentally achieved a high generation rate.

This work is supported by National Natural Science Foundation of China (Grant No. 61771439, 61501414, 61702469, 61602045, 61531003), National Cryptography Development Fund (Grant No. MMJJ20170120), Sichuan Youth Science and Technology Foundation(Grant No. 2017JQ0045), Foundation of Science and Technology on Communication Security Laboratory (Grant No. 6142103040105), China Postdoctoral Science Foundation (Grant No. 2018M630116).

\appendix

\section{Model of coarse-grained homodyne measurement}

\begin{figure}[b]
\centering
\subfigure[ ]{ \label{fig:subfig:a} 
\includegraphics[width=0.32 \textwidth]{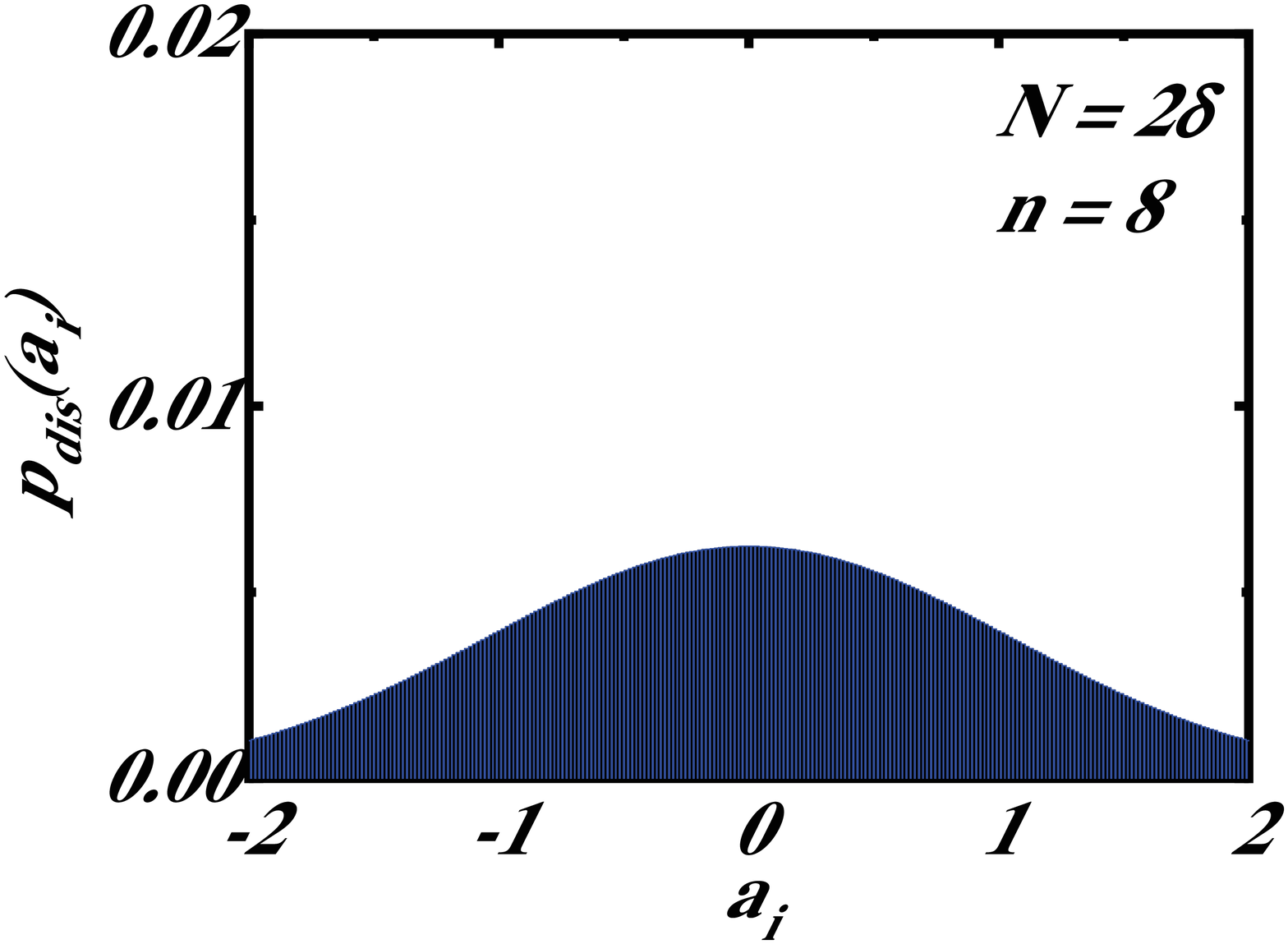}}
\hspace{1in}
\subfigure[ ]{ \label{fig:subfig:b} 
\includegraphics[width=0.32 \textwidth]{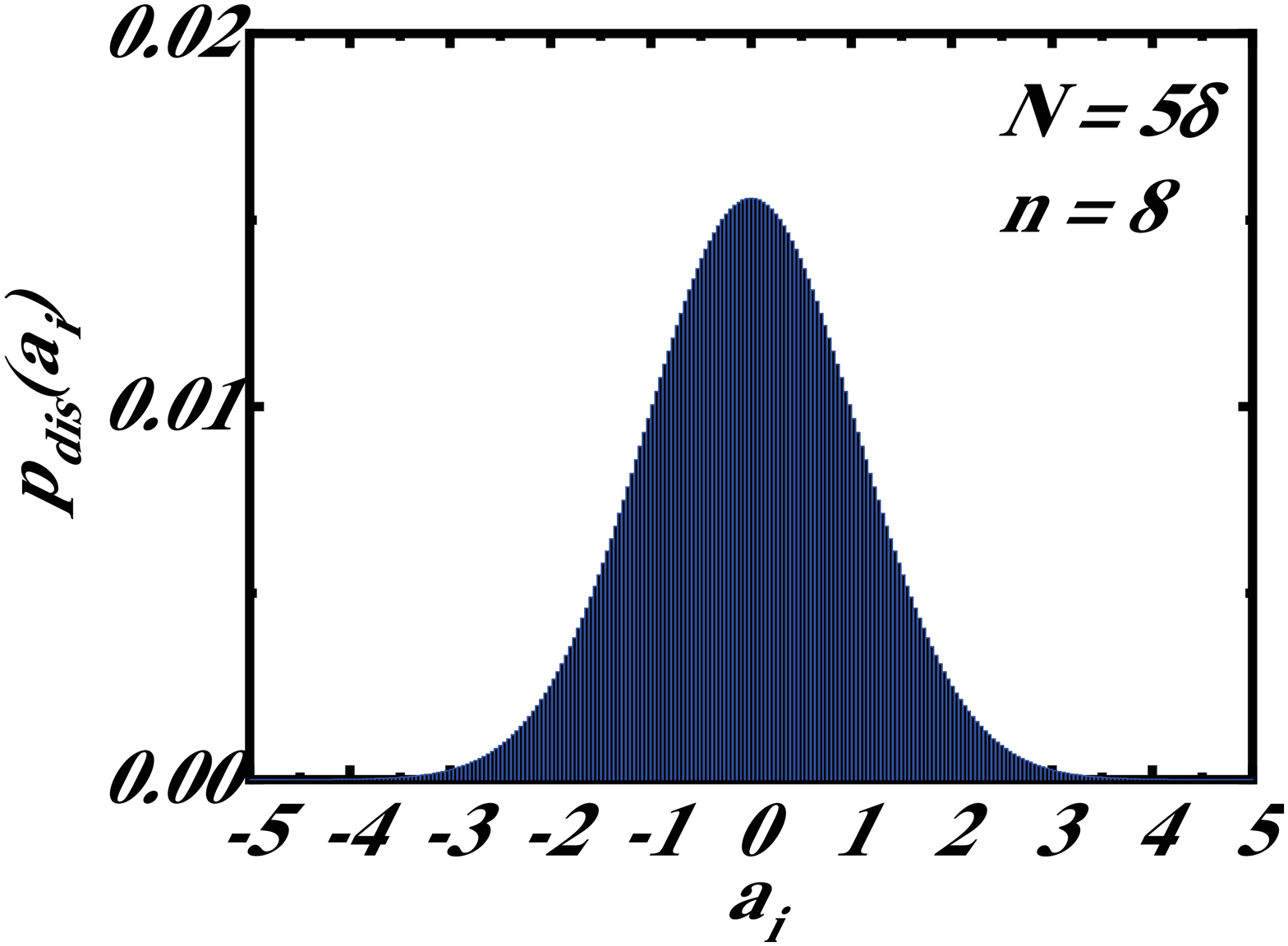}}
\hspace{1in}
\subfigure[ ]{ \label{fig:subfig:c} 
\includegraphics[width= 0.32 \textwidth]{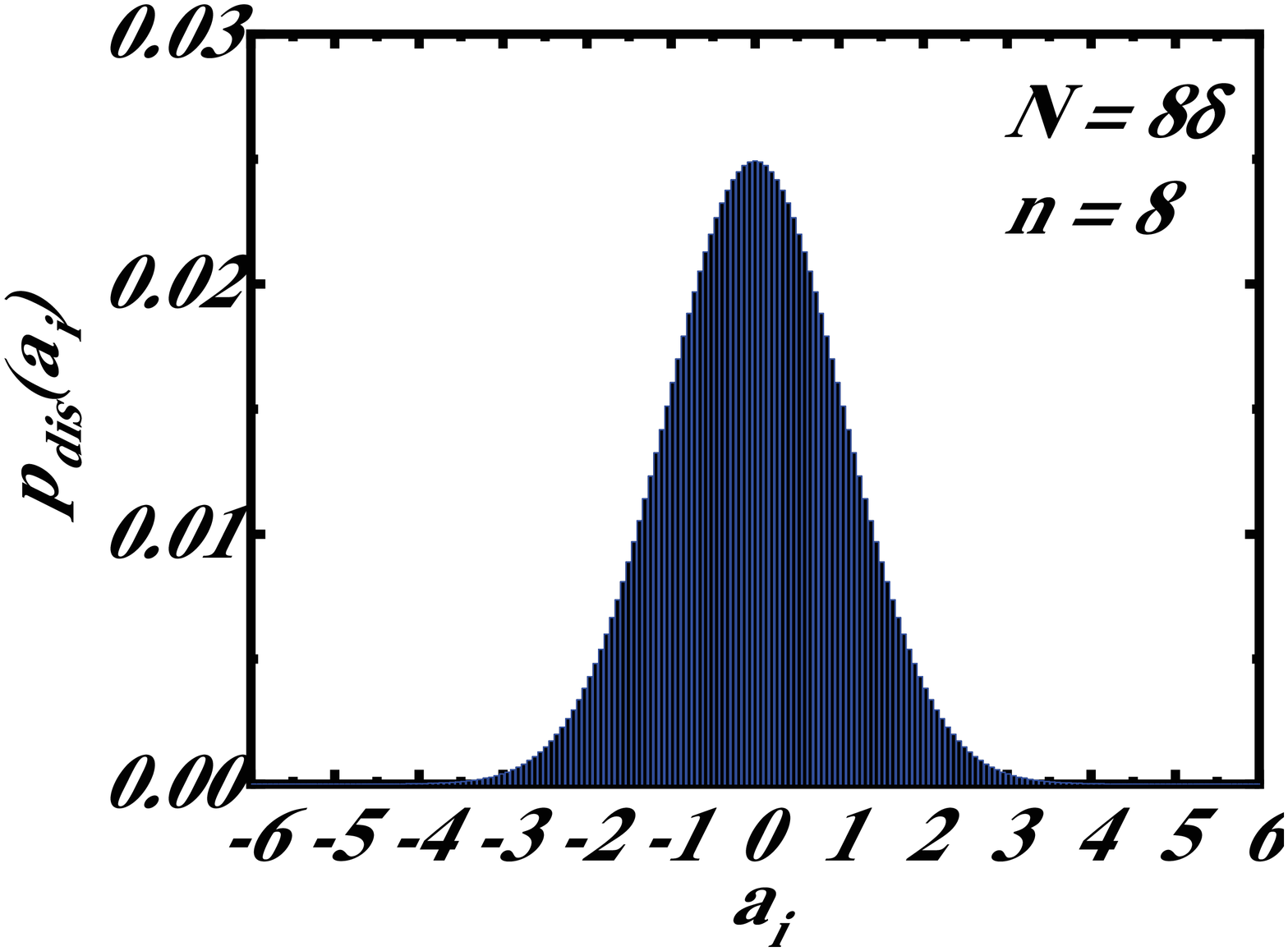}}
\hspace{1in}
\subfigure[ ]{ \label{fig:subfig:d} 
\includegraphics[width= 0.32 \textwidth]{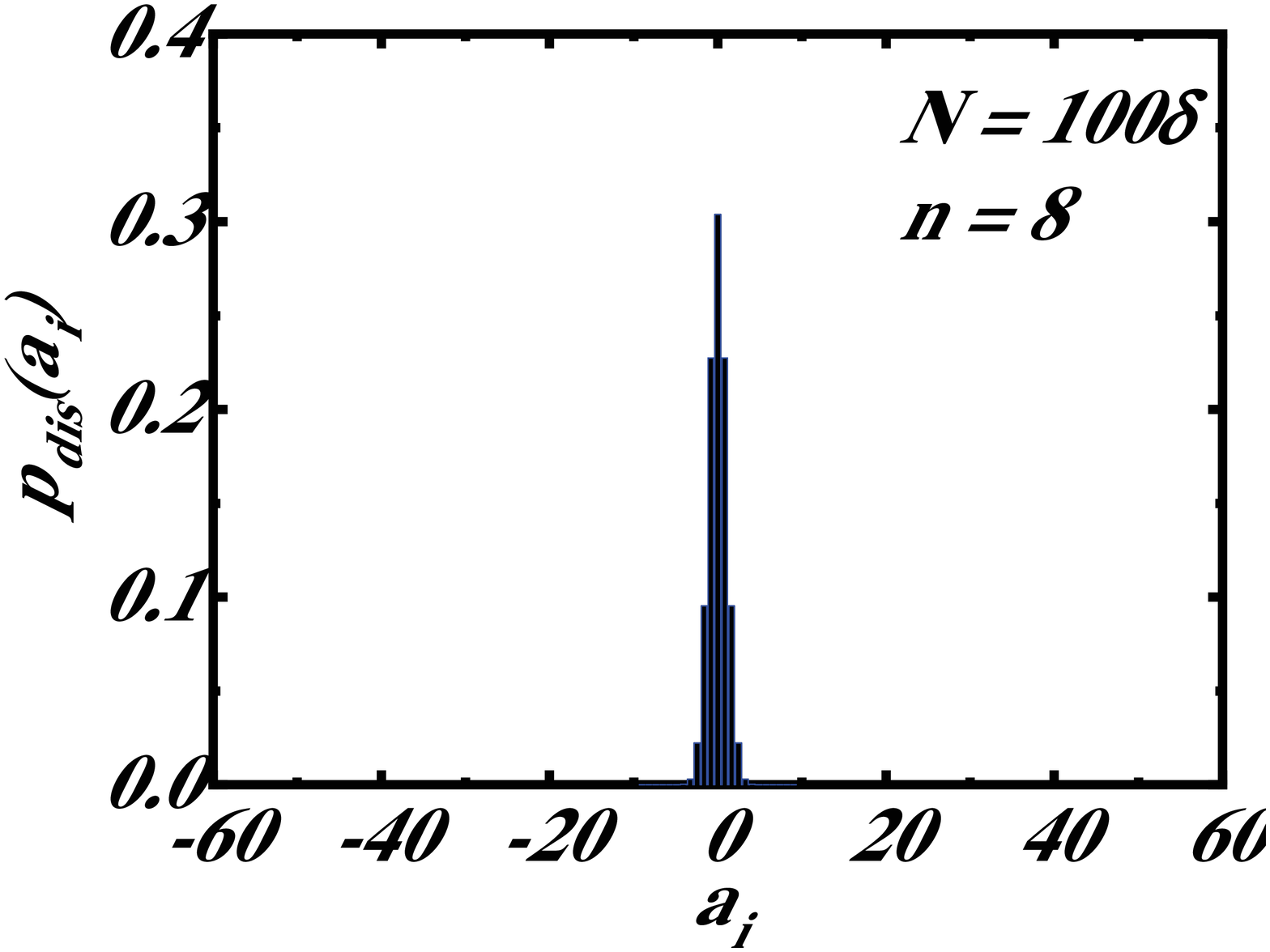}}
\caption{(color online) Simulation results of the probability distribution  $p_{dis}(a_i)$ of Alice's measurement results $a_i$ with different sampling range $N=2\sigma,\ 5\sigma,\ 8\sigma$ and $100\sigma$ by Eq. (A4) under Gaussian assumption. The resolution is fixed to be $n=8$ and the excess noise is chosen to be $\varepsilon=0.1$.}
\label{fig:PDF_vs_N} 
\end{figure}

\begin{figure}[t]
\centering
\subfigure[ ]{ \label{fig:subfig:a} 
\includegraphics[width=0.32 \textwidth]{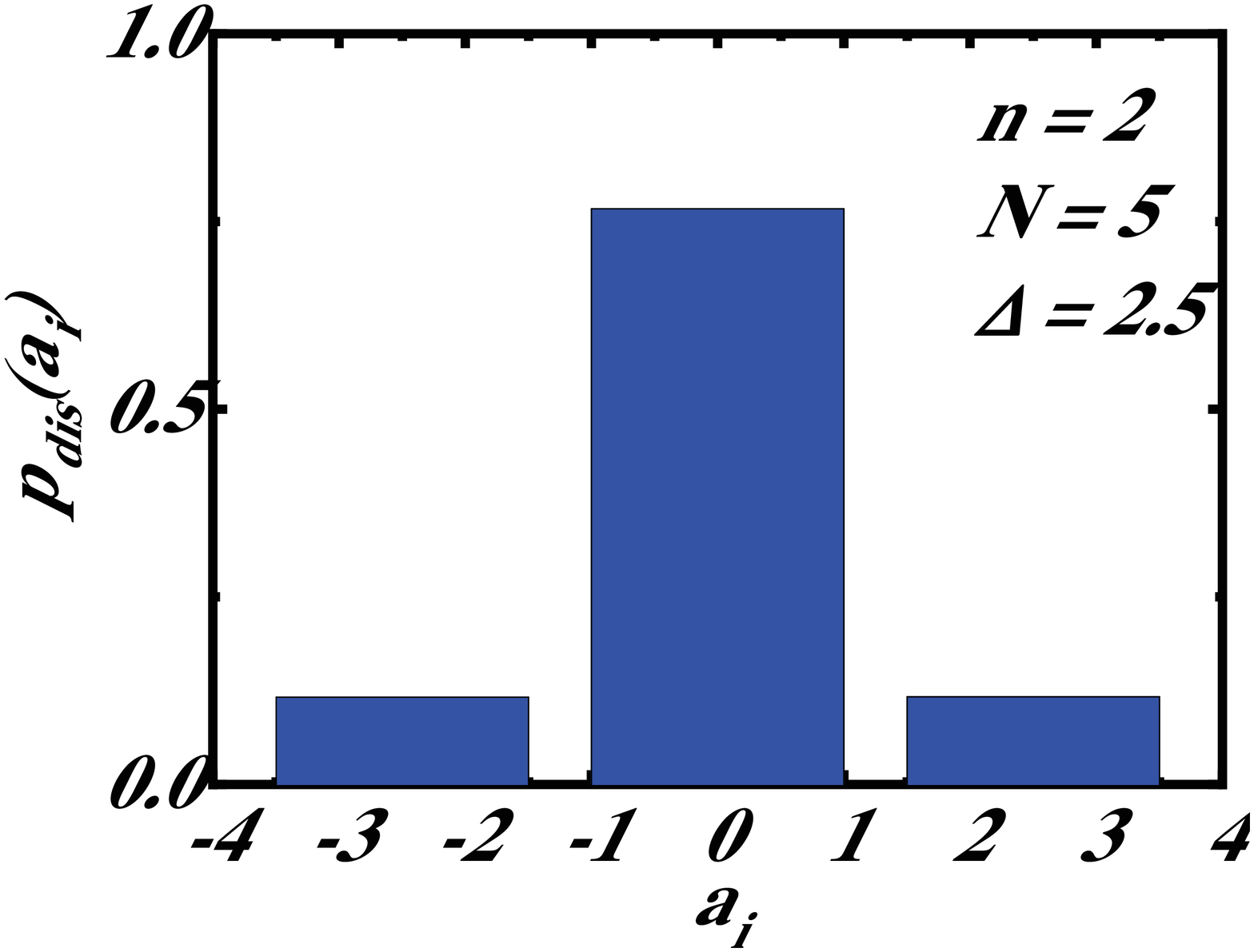}}
\hspace{1in}
\subfigure[ ]{ \label{fig:subfig:b} 
\includegraphics[width=0.32 \textwidth]{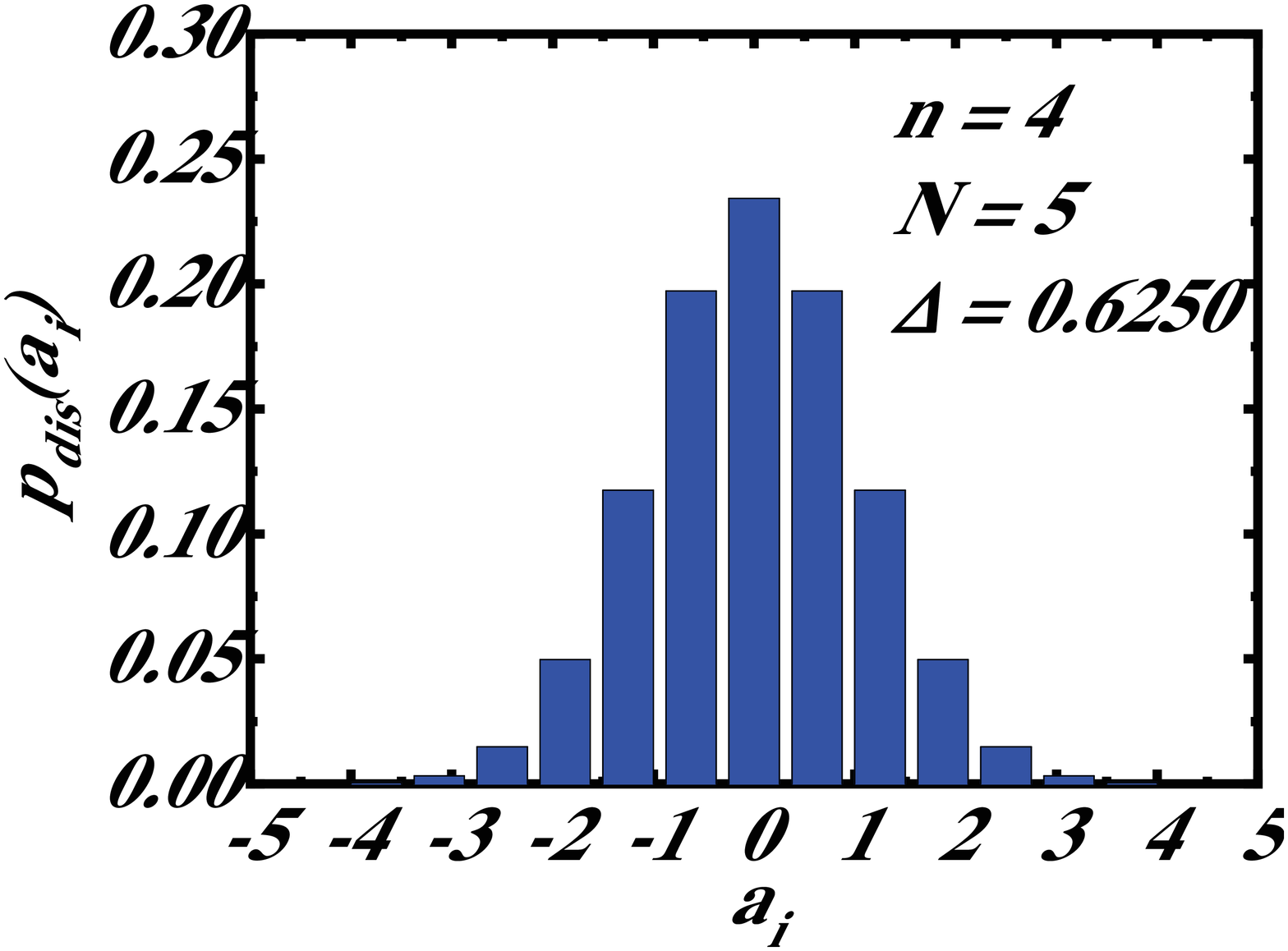}}
\hspace{1in}
\subfigure[ ]{ \label{fig:subfig:c} 
\includegraphics[width= 0.32 \textwidth]{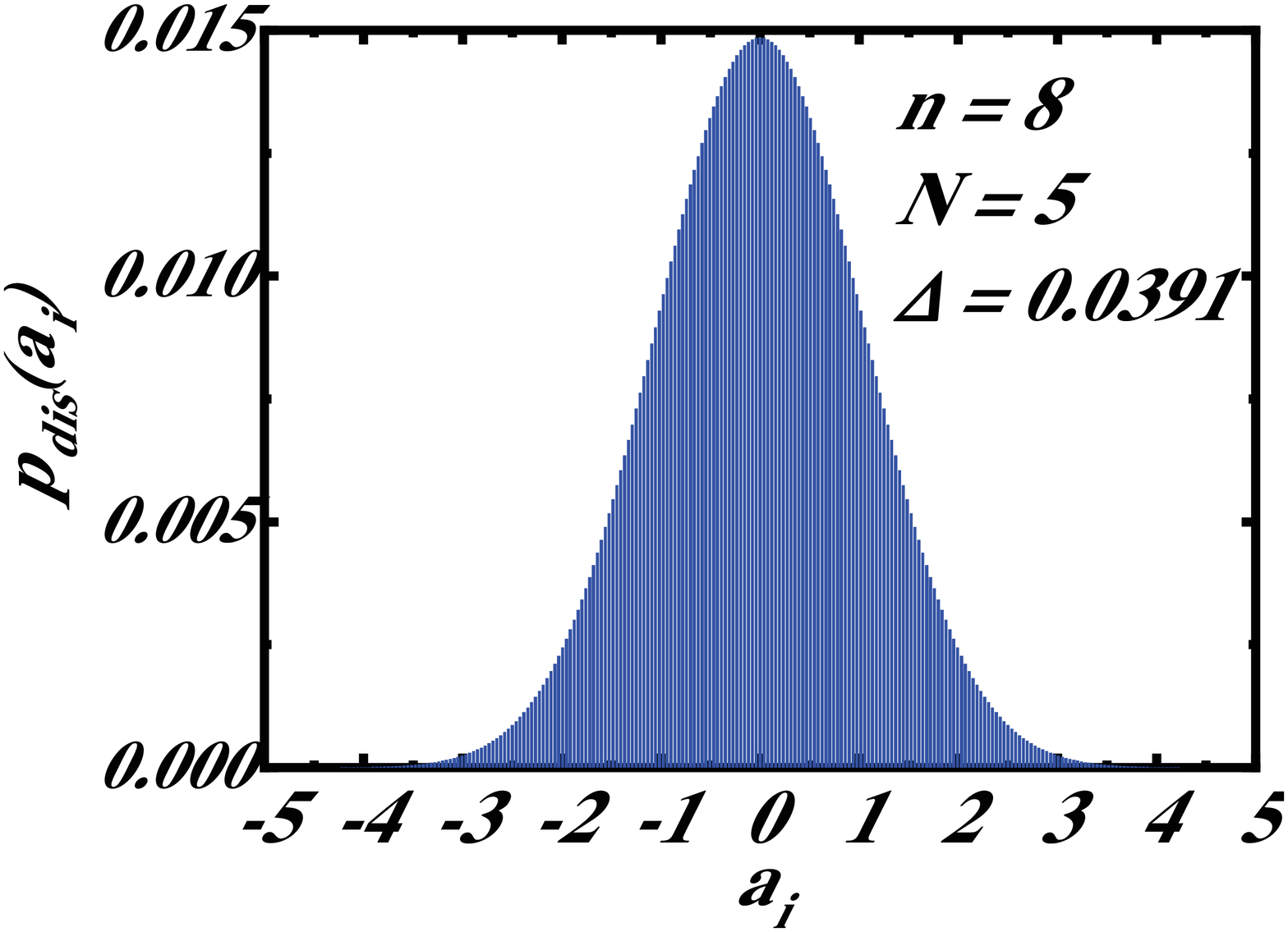}}
\hspace{1in}
\subfigure[ ]{ \label{fig:subfig:d} 
\includegraphics[width= 0.32 \textwidth]{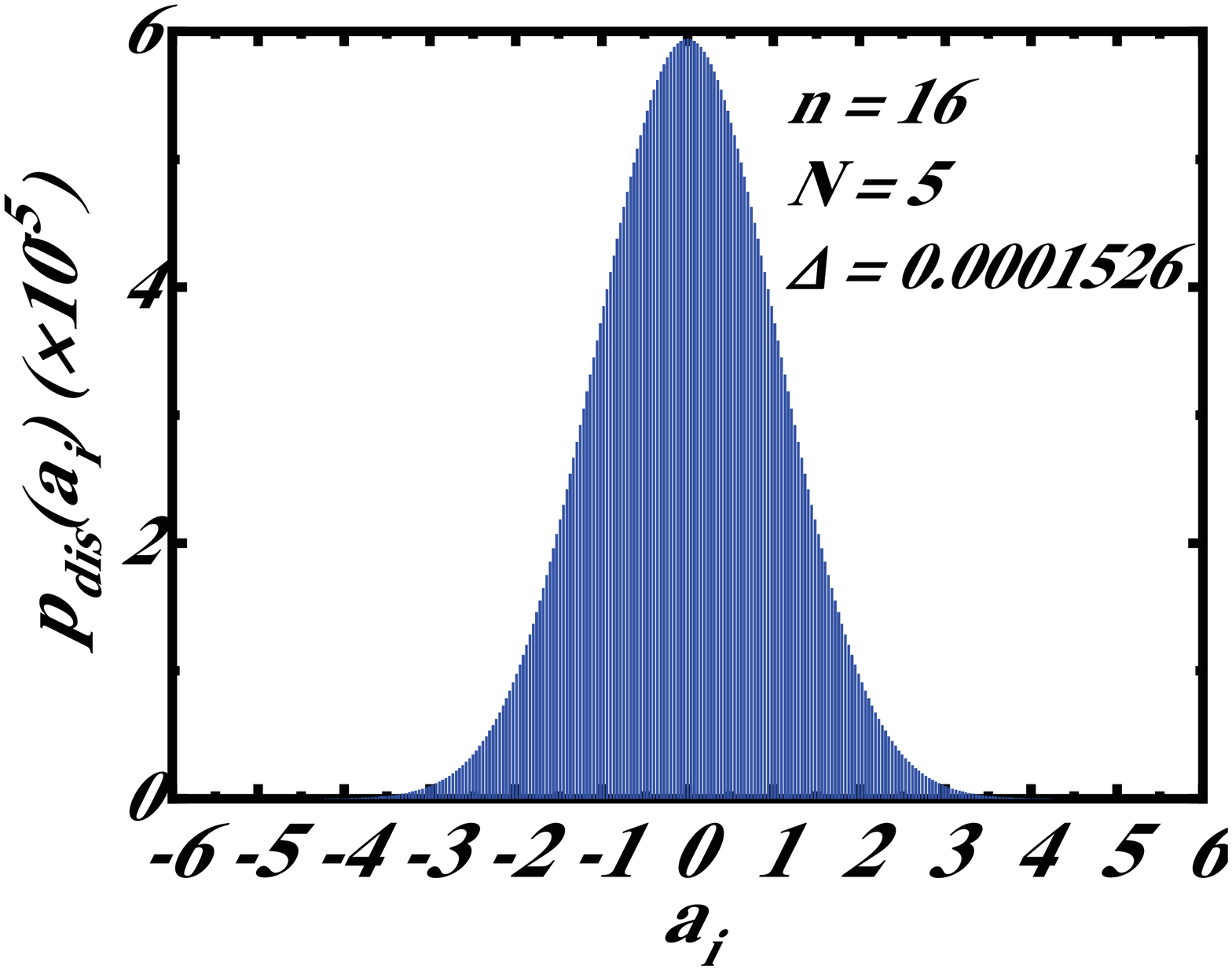}}
\caption{(color online) Simulation results of the probability distribution $p_{dis}(a_i)$  of Alice's measurement results $a_i$ with different resolution $n=2,\ 4,\ 8$ and $16$ by Eq. (A4) under Gaussian assumption. The sampling range is fixed to $N=5$ and the excess noise is chosen to be $\varepsilon  = 0.1$.}
\label{fig:Fig-PDF_vs_n} 
\end{figure}

We model Alice's real-life coarse-grained homodyne measurement process into two steps. Firstly, Alice uses an ideal homodyne detector to measure the quadrature of input states with continuous output $a$ following probability distribution $p(a)$, which cannot be read. Secondly, Alice digitizes the continuous variable $a$ into $n$ digitized bits $a_i$ following probability distribution $p_{dis}(a_i)$ by an ADC with sampling range $N$ and resolution $n$~\cite{QRNG_Vac4,QRNG_SDI2}, which is the actual output of a real-life homodyne detector. To be precise, Alice digitizes the data $a$ between $[ - N + \Delta /2,N - \Delta 3/2]$ into  $2^n$ equal intervals with bin width $\Delta  = N/{2^{n - 1}}$. The range is chosen so that the central bin is centered at zero. The central value of each interval represents the digitized results, while for data smaller than $- N$ or greater than $N$ it will be the smallest and the largest digitized value. Then, one has,
\begin{equation}
{a_i} = \left\{ {\begin{array}{*{20}{c}}
{{a_{ - {2^{n - 1}}}} \equiv  - N}&{a <  - N + \Delta /2}\\
{{a_{ - {2^{n - 1}} + 1}} \equiv  - N + \Delta }&{ - N + \Delta /2 \le a <  - N + 3\Delta /2}\\
{{a_{ - {2^{n - 1}} + 2}} \equiv  - N + 2\Delta }&{ - N + 3\Delta /2 \le a <  - N + 5\Delta /2}\\
 \vdots & \vdots \\
{{a_0} \equiv 0}&{ - \Delta /2 \le a < \Delta /2}\\
{{a_1} \equiv \Delta }&{\Delta /2 \le a < 3\Delta /2}\\
 \vdots & \vdots \\
{{a_{{2^{n - 1}} - 3}} \equiv N - 3\Delta }&{N - 7\Delta /2 \le a < N - 5\Delta /2}\\
{{a_{{2^{n - 1}} - 2}} \equiv N - 2\Delta }&{N - 5\Delta /2 \le a < N - 3\Delta /2}\\
{{a_{{2^{n - 1}} - 1}} \equiv N - \Delta }&{a \ge N - 3\Delta /2}
\end{array}} \right.,
\end{equation}
where $i \in \left\{ { - {2^{n - 1}}, \ldots ,{2^{n - 1}} - 1} \right\}$. The probability distribution $p_{dis}(a_i)$ reads,
\begin{equation}
p_{dis}({a_i}) = \left\{ \begin{array}{l}
\begin{array}{*{20}{c}}
{\mathop \smallint \limits_{ - \infty }^{ - N + \Delta /2} p\left( a \right)da}&{i = {i_{\min }}}
\end{array}\\
\begin{array}{*{20}{c}}
{\mathop \smallint \limits_{{a_i} - \Delta /2}^{{a_i} + \Delta /2} p\left( a \right)da}&{{i_{\min }} < i < {i_{\max }}}
\end{array}\\
\begin{array}{*{20}{c}}
{\mathop \smallint \limits_{N - 3\Delta /2}^{ + \infty } p\left( a \right)da}&{i = {i_{\max }}}
\end{array}
\end{array} \right.
\end{equation}
Suppose $a$  follows Gaussian distribution with variance ${\sigma ^2} = 1 + \varepsilon$  and null mean value,
\begin{equation}
p\left( a \right) = \frac{1}{{\sqrt {2\pi } \sigma }}exp\left( { - \frac{{{a^2}}}{{2{\sigma ^2}}}} \right),
\end{equation}
which is a typical result in vacuum fluctuation measurement. Then,
\begin{equation}
p_{dis}^G({a_i}) = \left\{ \begin{array}{l}
\begin{array}{*{20}{c}}
{\frac{1}{2}erfc(\frac{{N - 0.5\Delta }}{{\sqrt 2 \sigma }}),}&{i = {i_{\min }}}
\end{array}\\
\begin{array}{*{20}{c}}
{\frac{1}{2}erf(\frac{{i + 0.5}}{{\sqrt 2 \sigma }}\Delta ) - \frac{1}{2}erf(\frac{{i - 0.5}}{{\sqrt 2 \sigma }}\Delta ),}&{{i_{\min }} < i < {i_{\max }}}
\end{array}\\
\begin{array}{*{20}{c}}
{\frac{1}{2}erfc(\frac{{N - 1.5\Delta }}{{\sqrt 2 \sigma }}),}&{i = {i_{\max }}}
\end{array}
\end{array} \right.
\end{equation}

Simulation results of the probability distribution $p_{dis}(a_i)$ with different sampling range and fixed resolution are shown in Fig.~\ref{fig:PDF_vs_N}. It is shown that $a_i$  becomes more predictable when $N$ is too  small (or large) due to the oversaturated (or unoccupied) measurement results, which will reduce the extractable randomness.

Simulation results of the probability distribution $p_{dis}(a_i)$ with different resolution and fixed sampling range are shown in Fig.~\ref{fig:Fig-PDF_vs_n}. If it is small, most measurement results lie in central bins, which will reduce the Shannon entropy dramatically. The larger the resolution, the better the extractable randomness.

\section{${\rm{S}}\left( {{a_i}:E} \right) \le S\left( {a:E} \right)$}
Suppose Alice's measurement results of an ideal and a coarse-grained  homodyne measurement on $X$ quadratures of quantum state $\rho_A$ are $a$ and $a_i$ (as in Appendix A), respectively, and denote the corresponding quantum state hold by Eve is ${\rho _{E|a}}$ and ${\rho _{E|a_i}}$, respectively. Then one has
\begin{equation}
{\rho _E} = \int {p\left( a \right){\rho _{E|a}}{\rm{da}}},
\end{equation}
and
\begin{equation}
{\rho _{E|{a_i}}} = \left\{ {\begin{array}{*{20}{c}}
{\int\limits_{ - \infty }^{ - N + \frac{1}{2}\Delta } {\frac{{p\left( a \right)}}{{p_{dis}\left( {{a_i}} \right)}}{\rho _{E|a}}} da}&{i = {i_{\min }}}\\
{\int\limits_{{a_i} - \frac{1}{2}\Delta }^{{a_i} + \frac{1}{2}\Delta } {\frac{{p\left( a \right)}}{{p_{dis}\left( {{a_i}} \right)}}{\rho _{E|a}}} da}&{{i_{\min }} < i < {i_{\max }}}\\
{\int\limits_{N - \frac{3}{2}\Delta }^{ + \infty } {\frac{{p\left( a \right)}}{{p_{dis}\left( {{a_i}} \right)}}{\rho _{E|a}}} da}&{i = {i_{\max }}}
\end{array}} \right.
\end{equation}
Note that  $p(a)$ is unknown to Alice. Thus, the total system is assumed to be
\begin{equation}
\rho _{_{AE}}^{dis} = \sum\limits_{i = {i_{\min }}}^{i = {i_{\max }}} {p_{dis}\left( {{a_i}} \right)\left| {{a_i}} \right\rangle \left\langle {{a_i}} \right| \otimes {\rho _{E|{a_i}}}}
\end{equation}
One can easily verify that the overall state of Eve is the same as the ideal case,
\begin{equation}
\rho _E^{dis} = tr(\rho _{_{AE}}^{dis}) = tr({\rho _{AE}}) = {\rho _E}.
\end{equation}
Using Holevo's bound, one has
\begin{equation}
I\left( {{a_i}:E} \right) \le S\left( {{a_i}:E} \right) = S\left( {\rho _E^{dis}} \right) - \sum\limits_{i = {i_{\min }}}^{i = {i_{\max }}} {p_{dis}\left( {{a_i}} \right)S\left( {{\rho _{E|{a_i}}}} \right)}.
\end{equation}
Using the concavity of the von Neumann entropy,
\begin{equation}
S\left( {{\rho _{E|{a_i}}}} \right) = S\left( {\int\limits_{{a_i} - \frac{1}{2}\Delta }^{{a_i} + \frac{1}{2}\Delta } {\frac{{p\left( a \right)}}{{p_{dis}\left( {{a_i}} \right)}}{\rho _{E|a}}} da} \right) \ge \int\limits_{{a_i} - \frac{1}{2}\Delta }^{{a_i} + \frac{1}{2}\Delta } {\frac{{p\left( a \right)}}{{p_{dis}\left( {{a_i}} \right)}}S({\rho _{E|a}})} da,
\end{equation}
(similar for $i=i_{\min}$ and $i_{\max})$. Then, one can get
\begin{equation}
\begin{array}{l}
I\left( {{a_i}:E} \right) \le S\left( {{a_i}:E} \right) = S\left( {\rho _E^{dis}} \right) - \sum\limits_{i = {i_{\min }}}^{i = {i_{\max }}} {p_{dis}\left( {{a_i}} \right)S\left( {{\rho _{E|{a_i}}}} \right)} \\
 \le S\left( {{\rho _E}} \right) - \int\limits_{ - \infty }^\infty  {p\left( a \right)S\left( {{\rho _{E|a}}} \right)} da = S\left( {E:a} \right)
\end{array}
\end{equation}
Qualitatively speaking, the digitization process can have a corresponding quantum operation acting only on Alice's side, while the operation conducting only on one part of the state can not increase the mutual information between two parties.

\section{Effects of digitization on estimation of CM}

The upper bound of $S(\rho_A)$ for $\rho_A$ can be calculated by considering it is a Gaussian state with the same CM using Gaussian extremality theorem. The real CM for $\rho_A$  should be estimated through Alice's ideal detection result $a$, which is continuous, noted by
\begin{equation}
{\gamma _A} = \left( {\begin{array}{*{20}{c}}
{{V_x}}&c\\
c&{{V_p}}
\end{array}} \right),
\end{equation}
with symplectic eigenvalue $\lambda  = \sqrt {\det ({\gamma _A})}  = \sqrt {{V_x}{V_p} - {c^2}} $, where $V_x$, $V_p$ and $c$ are real numbers. The von Neumann entropy of a Gaussian state with CM  $\gamma _A$ is
\begin{equation}
S(\rho _A^G) = \frac{{\lambda  + 1}}{2}{\log _2}\frac{{\lambda  + 1}}{2} - \frac{{\lambda  - 1}}{2}{\log _2}\frac{{\lambda  - 1}}{2} \ge S({\rho _A}).
\end{equation}
To estimate the upper bound of  $S(\rho^G_A)$, one only needs to upper bound the symplectic eigenvalue, which is equivalent to upper bound $V_x$  and $V_p$, lower bound $c^2$.

\begin{figure}[b]
\begin{center}
\includegraphics[width=0.6\textwidth]{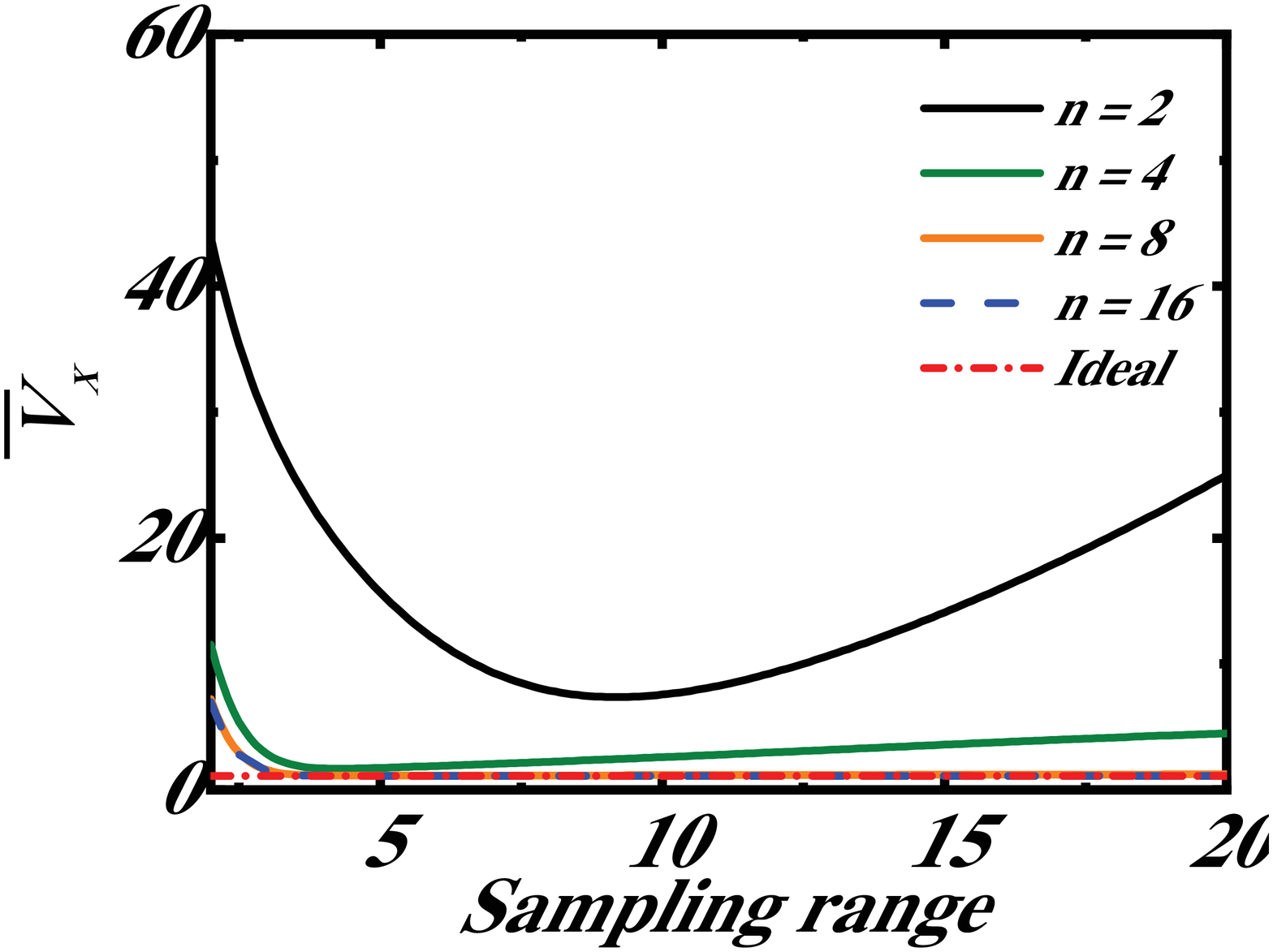}
\end{center}
\caption{(color online) The simulation results for upper bound of quadratures variance ${\overline V_x} $ (${\overline V_p}$) for $\rho_A$ based on Alice's digitized measurement results as a function of sampling range N with resolution $n=2,\ 4,\ 8,\ 16$, respectively. The ideal case corresponds to the estimated value of $V_x$ when $N \to \infty ,n \to \infty$. The excess noise is chosen to be $\varepsilon  = 0.1$.}
\label{fig:Vx_vs_R_n}
\end{figure}

\begin{figure}[t]
\begin{center}
\includegraphics[width=0.6\textwidth]{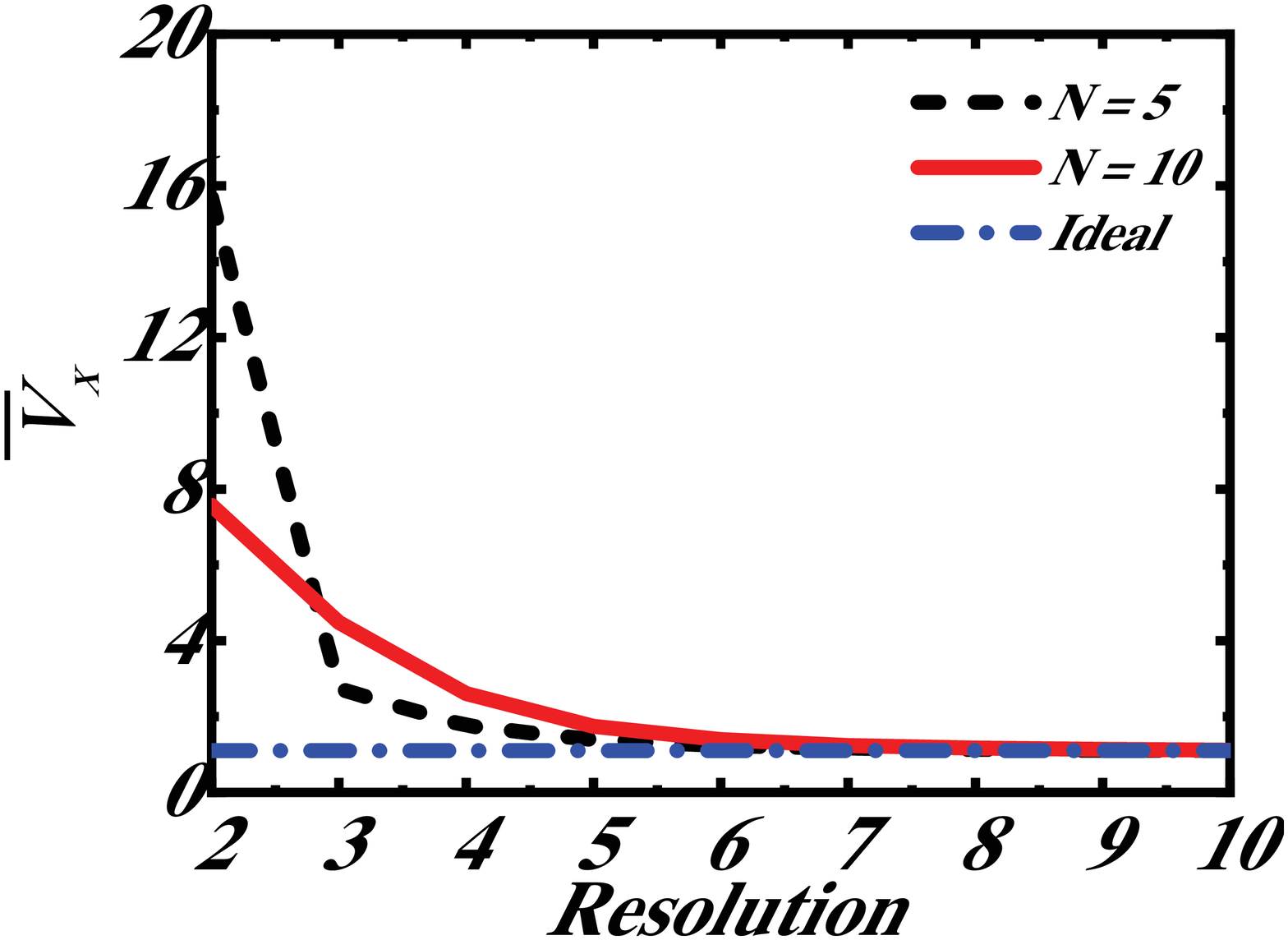}
\end{center}
\caption{(color online) The simulation results for upper bound of quadratures variance ${\overline V_x}$ for $\rho_A$ based on Alice's digitized measurement results as a function of resolution $n$ and range $N=5, 10$, respectively. The ideal case corresponds to the estimated value of $V_x$ when $N \to \infty ,n \to \infty$. The excess noise is chosen to be $\varepsilon  = 0.1$.}
\label{fig:Vx_vs_n_R}
\end{figure}

In ideal digitization with infinite sampling range, the upper bound of $V_x$($V_p$) is easy to get, since each unknown $a$ is upper and lower bounded by its digitization interval, \emph{i.e.}, $a_i - \frac{1}{2}\Delta \le a \le a_i + \frac{1}{2} \Delta $. Therefore, for $a_i \le 0$ we use $a_i - \frac{1}{2} \Delta$, and for $a_i>0$ we use $a_i + \frac{1}{2}\Delta$ to calculate $V_x$, which is an upper bound. However, in a real system, the digitization has finite sampling range. For asymptotic case, an extra energy test is needed for the most rigorous proof. For a approximate but more practical solution, we can set a relatively large bound $\left[-a_{\lim},a_{\lim} \right]$, in which the probability of the case $a \notin \left[-a_{\lim},a_{\lim} \right]$ is negligible. For example, for a Gaussian distributed $a$ with variance $\sigma^2$, if ${a_{\lim }} = 10\sigma$, then  $Pr\left[a \notin \left[-a_{\lim},a_{\lim}\right]\right] < 1.5 \times {10^{ - 23}}$; and if ${a_{\lim }} = 100\sigma$, then $Pr\left[a \notin \left[-a_{\lim},a_{\lim}\right]\right] < 5.5 \times {10^{ - 89}}$. Considering more general cases, the measurement results are in the interval $\left[ {{a_{\min }},{a_{\max }}} \right]$ with non-zero mean $\bar a$. We use ${a_i}$ ($i \in \left[ {{i_{\min }},0} \right)$) to represent the results which are smaller than the mean value, and ${a_i}$ ($i \in \left( {0,{i_{\max }}} \right]$) denotes the results which are larger than the mean. The $V_x$ is upper bounded by the following quantity with high confidence level,
\begin{eqnarray}
{\overline V_x} &=& p_{dis}\left( {{a_{{i_{\min }}}}} \right)(a_{\min }-\bar a)^2 + p_{dis}\left( {{a_{{i_{\max }}}}} \right)(a_{\max }-\bar a)^2  \\ \nonumber
 &+& \sum\limits_{i = {i_{\min }} + 1}^0 {p_{dis}\left( {{a_i}} \right){{\left( {{a_i} - \bar a - \frac{1}{2}\Delta } \right)}^2}}  + \sum\limits_{i =  + 1}^{{i_{\max }} - 1} {p_{dis}\left( {{a_i}} \right){{\left( {{a_i} - \bar a + \frac{1}{2}\Delta } \right)}^2}}.
\end{eqnarray}

Similarly, one can upper bound the variance of $P$ quadrature with the same methods, ${\overline V_p}  \ge {V_p}$. For simplicity, we set $c=0$ to upper bound Eve's information. Finally, the upper bound of the symplectic eigenvalue for $\rho_A$ is $\overline \lambda   = \sqrt {{\overline V_x} {\overline V_p}} $, and
\begin{equation}
S({\rho _A}) \le \frac{{\overline \lambda   + 1}}{2}{\log _2}\frac{{\overline \lambda   + 1}}{2} - \frac{{\overline \lambda   - 1}}{2}{\log _2}\frac{{\overline \lambda   - 1}}{2}.
\end{equation}

The effects of sampling range $N$ and resolution $n$ on estimated upper bound of quadrature variance is shown in Fig.~\ref{fig:Vx_vs_R_n} and Fig.~\ref{fig:Vx_vs_n_R}, respectively. Generally, one can get a tighter upper bound ${\overline V_x}$ by increasing $n$. When $N<3\sigma$, one will overestimate the variance of $a$ based on measurement results $a_i$, which will significantly overestimate Eve's information.

For the finite-size case, we can choose a relatively large sampling range $N$, and set the bound of energy test within $\left[-N,N\right]$. And we further require that even if only one digitized data exceeds the energy test bound, the energy test for the whole block fails. Therefore, for all the data blocks that pass the energy test, the upper bound of the estimated $V_x$ can be got through the same strategy as above. For the data blocks fail the energy test, this round of the protocol aborts. One should note that, the sampling range should be carefully chosen, if it's too small, too many blocks will fail; if it's too large, the estimated parameter will be too pessimistic.\\

\end{document}